\documentclass[a4paper,fleqn,usenatbib]{mnras}
\usepackage[T1]{fontenc}
\usepackage{ae,aecompl}
\usepackage{graphicx}
\usepackage{footnote}
\usepackage[fleqn]{amsmath}
\usepackage[varg]{txfonts}
\usepackage{multirow}
\usepackage{subfig}
\newcommand\numberthis{\addtocounter{equation}{1}\tag{\theequation}}

\newcommand{\arXiv}{ArXiv}
\newcommand{\aec}{A\&C}

\title[New methods to constrain the radio transient rate]{New methods to constrain the radio transient rate:
results from\newline a survey of four fields with LOFAR}

\author[D. Carbone et al.]
{
D.~Carbone$^{1}$\thanks{E-mail: d.carbone@uva.nl}, A.~J.~van~der~Horst$^{1,2}$, R.~A.~M.~J.~Wijers$^{1}$, J.~D.~Swinbank$^{1, 3}$,\newauthor
A.~Rowlinson$^{1, 4, 13}$, J.~W.~Broderick$^{5, 6, 13}$, Y.~N.~Cendes$^{1}$, A.~J.~Stewart$^{5}$, M.~E.~Bell$^{4, 7}$,\newauthor
R.~P.~Breton$^{8}$, S.~Corbel$^{9, 10}$, J.~Eisl\"{o}ffel$^{11}$, R.~P.~Fender$^{5}$, J.-M.~Grie{\ss}meier$^{10, 12}$,\newauthor
J.~W.~T.~Hessels$^{1, 13}$, P.~Jonker$^{14, 15}$, M.~Kramer$^{8, 16}$, C.~J.~Law$^{17}$,  J.~C.~A.~Miller-Jones$^{18}$,\newauthor
M.~Pietka$^{5}$, L.~H.~A.~Scheers$^{1, 19}$, B.~W.~Stappers$^{8}$, J.~van~Leeuwen$^{1, 13}$, R.~Wijnands$^{1}$,\newauthor
M.~Wise$^{1, 13}$, P.~Zarka$^{20}$\\
$^{1}$ Anton Pannekoek Institute for Astronomy, University of Amsterdam, Postbus 94249, 1090 GE Amsterdam, The Netherlands \\
$^{2}$ Department of Physics, The George Washington University, 725 21$^{st}$ Street NW, Washington, DC 20052, USA \\
$^{3}$ Department of Astrophysical Sciences, Princeton University, Princeton, NJ 08544, USA \\
$^{4}$ CSIRO Australia Telescope National Facility, PO Box 76, Epping NSW 1710, Australia \\
$^{5}$ Oxford Astrophysics, Denys Wilkinson Building, Keble Road, Oxford OX1 3RH, UK \\
$^{6}$ Physics and Astronomy, University of Southampton, Highfield, Southampton SO17 1BJ, UK \\
$^{7}$ ARC Centre of Excellence for All-sky Astrophysics (CAASTRO), The University of Sydney, NSW 2006, Australia \\
$^{8}$ Jodrell Bank Centre for Astrophysics, School of Physics and Astronomy, The University of Manchester, Manchester M13 9PL, UK \\
$^{9}$ Laboratoire AIM (CEA/IRFU - CNRS/INSU - Universit\'e Paris Diderot), CEA DSM/IRFU/SAp, F-91191 Gif-sur-Yvette, France \\
$^{10}$ Station de Radioastronomie de Nan\c{c}ay, Observatoire de Paris, CNRS/INSU, USR 704 - Univ. Orl\'{e}ans, OSUC, 18330 Nan\c{c}ay, France \\
$^{11}$ Th\"{u}ringer Landessternwarte Tautenburg, Sternwarte 5, D-07778 Tautenburg,Germany  \\
$^{12}$ Laboratoire de Physique et Chimie de l'Environnement et de l'Espace, LPC2E UMR 7328 CNRS, 45071 Orl\'{e}ans, France \\
$^{13}$ ASTRON, The Netherlands Institute for Radio Astronomy, Postbus 2, 7990 AA Dwingeloo, The Netherlands \\
$^{14}$ SRON, Netherlands Institute for Space Research, Sorbonnelaan 2, 3584-CA, Utrecht, The Netherlands \\
$^{15}$ Department of Astrophysics/IMAPP, Radboud University Nijmegen, PO Box 9010, 6500 GL Nijmegen, The Netherlands \\
$^{16}$ Max-Planck-Institut f\"{u}r Radioastronomie, Auf dem H\"{u}gel 69, 53121 Bonn, Germany \\
$^{17}$ Department of Astronomy and Radio Astronomy Lab, University of California, Berkeley, CA, USA \\
$^{18}$ International Centre for Radio Astronomy Research, Curtin University, GPO Box U1987, Perth, WA 6845, Australia \\
$^{19}$ Centrum Wiskunde \& Informatica, P.O. Box 94079, 1090 GB Amsterdam, The Netherlands \\
$^{20}$ LESIA, Observatoire de Paris, CNRS, UPMC, Universit\'{e} Paris-Diderot, 5 place Jules Janssen, 92195 Meudon, France \\
}

\date{Accepted XXX. Received YYY; in original form ZZZ}
\pubyear{2015}

\begin{document}
\label{firstpage}
\pagerange{\pageref{firstpage}--\pageref{lastpage}}
\maketitle

\begin{abstract}
We report on the results of a search for radio transients between 115 and 190\,MHz with the LOw-Frequency
ARray (LOFAR). Four fields have been monitored with cadences between 15 minutes and several months.
A total of 151 images were obtained, giving a total survey area of 2275\,deg$^2$. We analysed our
data using standard LOFAR tools and searched for radio transients using the LOFAR Transients Pipeline
(TraP). No credible radio transient candidate has been detected; however, we are able to
set upper limits on the surface density of radio transient sources at low radio frequencies.
We also show that low-frequency radio surveys are more sensitive to steep-spectrum coherent transient sources than
GHz radio surveys.
We used two new statistical methods to determine the upper limits on the transient surface density.
One is free of assumptions on the flux distribution of the sources, while the other assumes a power-law
distribution in flux and sets more stringent constraints on the transient surface density. 
Both of these methods provide better constraints than the approach used in previous works.
The best value for the upper limit we can set for the transient surface density, using the method assuming
a power-law flux distribution,
is 1.3\,$\cdot$\,10$^{-3}$\,deg$^{-2}$ for transients brighter than 0.3\,Jy
with a time-scale of 15\,min, at a frequency of 150\,MHz.
We also calculated for the first time upper limits for the transient surface density for transients of different
time-scales.
We find that the results can differ by orders of magnitude from previously reported, simplified estimates.
\end{abstract}


\section{Introduction}
\label{sec:intro}

Radio emission often signals relatively exotic, non-thermally emitting sources, in which relativistic processes
play a significant role. Since these sources are often compact objects, they are also able to change their
brightness on short time-scales, and indeed they are often found to be variable or transient. Traditionally,
such variability was often found after discovery as a salient property of the source, but the advent of very
powerful wide-field telescopes combined with rapid data processing capabilities has given rise to time-domain
astronomy, in which the variability or transience of a source becomes the prime property by which it is
discovered. In X-ray and $\gamma$-ray astronomy this has been the case almost since the beginning,
and indeed most sources are variable at those high photon energies. In radio and optical, time-domain
astronomy in this sense is younger and still in an exploratory phase, with new discovery space being
opened up all the time.
Radio transient behaviour can probe a great variety of source types, over a wide range of observational
parameters, such as distance and time-scale.

So far, discoveries of new transient sources at radio wavelength have been sparse due to insufficient
sky coverage of surveys with adequate sensitivity and time resolution; these surveys have instead often
led to the discovery of time-dependent behaviour of previously known sources.
However, in the last few years, a new generation of wide-field
facilities has been available to sample the transient sky, creating the opportunity to discover sources as
transients by blind surveys also in radio.
The Allen Telescope Array
\citep[ATA;][]{ATA_Welch2009} carried out transient surveys from 2007 and 2011; currently, the Karl
G. Jansky Very Large Array \citep[VLA;][]{JVLA_Perley2011}, the Murchison Wide Field Array
\citep[MWA;][]{MWA_tingay2013},  the Long Wavelength Array \citep[LWA;][]{LWA_Ellingson2009},
and the LOw Frequency ARray \citep[LOFAR;][]{LOFAR_vanHaarlem2013} are operational. 
In addition, the Karoo Array Telescope
\citep[MeerKAT;][]{Meerkat_Booth2009} and the Australian Square-Kilometer Array Pathfinder
\citep[ASKAP;][]{ASKAP_Johnston2008} are both currently about to start collecting data. Another radio
facility that will perform transient searches will be the APERture Tile In Focus 
\citep[APERTIF;][]{apertif_vanLeeuwen2014} at the Westerbork Synthesis
Radio Telescope. These instruments cover a wide range of frequency 
and time scale, and thus each covers a different part of the large
parameter space of time domain astronomy.

At low radio frequencies, two types of emission mechanisms are expected to play a role: incoherent
and coherent emission. Incoherent emission is constrained to obey the blackbody limit, and
since radio frequencies are always low compared to the peak of a thermal spectrum, it depends on
frequency as $\nu^2$. This obviously favours high frequencies, and the only way to overcome this is to
make the source very big so it has large emitting area. But a source of very large size $R$ can only vary
on a slow time-scale $t_{\rm var}=R/c$, and so bright low-frequency transients are expected to be very
slow, changing on time-scales of months to years. Once an outburst becomes large enough to be optically
thin, the brightness can rise to lower frequency, but since an optically thin source is always fainter than 
an optically thick one of the same size and temperature, that does not help its detection at low frequencies.
For this reason, some recent works have suggested that searching for transients at low radio frequencies will
not produce interesting results, at least until 
we have the sensitivities of the Square Kilometer Array (SKA) \citep[e.g.,][]{incoherent_tr_rate_Metzger2015}.
On the other hand, there are several types of source that produce coherent emission. The importance of these
is best illustrated by looking at the data: if we plot known types of radio
source in a diagram of luminosity as a function of duration\,$\times$\,frequency, we do indeed find many
sources that obey the blackbody limit for relatively low luminosity and long duration
(see, e.g., Figure~5 of \citealt[][]{timescale_Tb_Pietka2015}).
However, an even larger part of the parameter space is filled with luminous, rapid transients that violate
the limit by very large amounts, many discovered relatively recently. These are all coherent emitters,
which can be very luminous and yet rapidly variable: most famously and classically radio pulsars,
which were discovered at 80\,MHz \citep[][]{Pulsars_Hewish1968}, and more recently 
Fast Radio Bursts \citep[FRBs; e.g.,][]{FRBs_Lorimer2007}.

For an informed opinion on the expected fluxes and durations of short transients, we have to understand
coherent emission processes. What we know from incoherent emission physics is that the intrinsic
brightness temperature of sources is likely limited to $10^{11-12}$\,K \citep[e.g.,][ where brightness
temperature is defined as the value of $T$ in the Rayleigh-Jeans formula that yields the correct flux
of the source]{Tb_limit_Kellermann1969, Tb_limit_Singal1986}.
Sources having brightness temperatures above about 10$^{12}$\,K 
must emit coherently, have their emission relativistically boosted, or both. 
However, we understand the processes underlying such sources very poorly and so in this paper we
shall take the approach of expecting a wide range of known and unknown types
of source, and exploring as much of parameter space as our experiment allows.
One important thing to note, that is particularly relevant here, is that most of those coherent
emitters for which we know the properties of the radio spectrum have quite steep spectra, typically
going as $\nu^{-2}$ or even $\nu^{-3}$ \citep[see e.g.][]{pulsars_Lorimer2012, Coherent_Melrose2009},
in contrast with a typical $\nu^{-0.8}$ for optically thin synchrotron emission.
This means that low-frequency instruments such as LOFAR may be intrinsically at an advantage to find
coherent emitters (in addition to having larger fields of view).
While known coherent transients have mostly been found in beam-formed searches
and last milliseconds to seconds, more recently fast transients have been discovered in low
frequency image plane surveys.
For example, the sources GCRT~J1745-3009 \citep[][]{GCRT1745_Hyman2005} and GCRT~J1746-2757
\citep[][]{GCRT1746_Hyman2002} were detected at 330\,MHz with the VLA, while GCRT~J1742-3001
\citep[][]{GCRT1742_Hyman2009} was discovered at 235\,MHz with the Giant Metrewave Radio Telescope.
These sources showed bright flares lasting from minutes to a few hours.
More recently, the low-frequency radio transient ILT~J225347+862146 \citep[][]{Adam_Stewart2015}
was discovered at 60\,MHz with LOFAR, lasting about 10 minutes.
The only significant population of transient  radio sources previously known in this duration range are relatively
nearby and low luminosity flare stars, having fluxes of about 1\,Jy at 1.2\,GHz \citep[][]{flarestars_Osten2006}.

In this work we will present our transient search results in a campaign of LOFAR observations of four different
fields. LOFAR is a new generation radio interferometer built in the Netherlands and other European
countries, operating at frequencies between 30 and 250\,MHz \citep[][]{LOFAR_vanHaarlem2013}.
There are very few instruments operating at these frequencies with good time, spectral
and angular resolution, making LOFAR a valuable facility to explore an unsearched region of parameter space.
It is the primary goal of our study to explore what new types of source might exist in this part of the frequency-time scale
space ($\sim100$\,MHz, minutes-months, above the synchrotron brightness temperature limit), encouraged by the
few finds that have been made there already.

Since we did not significantly detect any transient sources, we focus in this paper on setting as good and precise
limits as the data allow on unknown transients, after concluding that methods in previous studies could be improved.
To this end, we discuss three different methods to determine the upper limit on the transient surface density, i.e.,
the number of transients per square degree brighter than a certain flux. 
We also discuss the time-scales to which our survey is sensitive, i.e., what the transient surface density is as a
function of time-scale for the transient sources that are searched for. We lay out the procedure for calculating this
duration dependence of the transient surface density, showing that this can vary significantly depending on the
time-scales that are probed in the survey. 
We describe the observational setup and data reduction in Section~\ref{sec:obs and data analysis}
and our results in Section~\ref{sec:results}. In Section~\ref{sec:methods} we
describe the statistical methods we used to derive our limits on the transient surface density, and determine the
transient surface density at different time-scales. We compare our results with the literature and discuss their implications
in Section~\ref{sec:discussion}, and summarise in Section~\ref{sec:conclusions}.

\begin{table*}
\begin{center}
\begin{tabular}{|c|c|c|c|c|c|c|c|c|}
\hline
\hline
\multirow{2}{*}{Field} 	&\multirow{2}{*}{RA (J2000)}	& \multirow{2}{*}{Dec (J2000)}	& \multirow{2}{*}{Galactic latitude (\textit{b})}&
\multirow{2}{*}{Galactic longitude (\textit{l})}	& Total number		& Number of		& Noise ratio	& Ellipticity 	\\
					& 	&						&									
&								& of images 		& images used 		& threshold	&  threshold	\\
\hline
MD03 	& 08$^\textrm{h}$42$^\textrm{m}$22$^\textrm{s}$	& +44$^{\circ}$19\arcmin00\arcsec	& 37.94$^{\circ}$	& 176.35$^{\circ}$	& 420	& 327	& 47.1	& 1.78	\\
MD05	& 10$^\textrm{h}$47$^\textrm{m}$40$^\textrm{s}$	& +58$^{\circ}$05\arcmin00\arcsec	& 52.24$^{\circ}$	& 149.29$^{\circ}$	& 420	& 156	& 62.5	& 1.67	\\
MD06	& 12$^\textrm{h}$20$^\textrm{m}$00$^\textrm{s}$	& +47$^{\circ}$07\arcmin00\arcsec	& 69.08$^{\circ}$	& 138.04$^{\circ}$	& 186 	& 149	& 59.9	& 1.34	\\
MD07	& 14$^\textrm{h}$14$^\textrm{m}$49$^\textrm{s}$	& +53$^{\circ}$05\arcmin00\arcsec	& 59.79$^{\circ}$	& 97.80$^{\circ}$	& 186	& 177	& 119.8	& 1.32	\\
\hline
\hline
\end{tabular}
\caption{Summary of the four fields we have observed. We provide the celestial coordinates and the number of images used
in our dataset before and after the quality control check.
The noise ratio and ellipticity thresholds used in the image quality control process (see Section~\ref{sec:results}) are determined
from the ratio between the measured and theoretical noise levels, and the ratio between the major and minor axis FWHMs of the
synthesized beam, respectively.  
Note that many images of MD05 have been rejected; this is because there was a misalignment of the station beam in the
first sessions which was solved after run 5.}
\label{tab:obs fields}
\end{center}
\end{table*}

\section{Observations \& Data analysis}
\label{sec:obs and data analysis}

\subsection{Observations}
\label{sec:observations}

We have observed four fields which form part of the PanSTARRS Medium Deep Field survey \citep[][]{MDfields_Tonry2012}.
We chose to monitor MD03, MD05, MD06 and MD07 because they can be observed at high elevation angles with LOFAR,
where the sensitivity is optimised. Moreover, although beyond the scope of this paper, these fields are potentially very
interesting for multi-wavelength studies. Field coordinates can be found in Table~\ref{tab:obs fields}.

A detailed description of LOFAR can be found in \citet{LOFAR_vanHaarlem2013}.
Our observations made use of 37 Dutch stations, which span a range of projected baselines between about 30\,m to
just under 80\,km. 
We used the LOFAR High Band Antennae (HBA), which cover a frequency range 110-250\,MHz, although we observed
from 115-190\,MHz only. The array configuration was HBA DUAL INNER, which means that
only the inner half of the antennae of the remote stations are correlated.
This means we reduced the remote HBA stations to the same size and the same field of view as the core HBA stations.
Although this implies a small loss in sensitivity, commissioning tests have shown it to increase image quality.
The primary beam full-width at half-maximum at 140\,MHz is about 4.1\,degrees in the array configuration we used.
Each snapshot of each field covers a total field of view of 27.92\,deg$^2$ in a circular region around the phase centre.
However, only the inner 15.48\,deg$^2$ have been searched for transients, as the outer part of the field of view is much
noisier and the flux calibration less reliable. 

Observations were carried out in 2013 March--August.
We started observing MD03 and MD05 only, for 2 hours every 2 weeks, simultaneously.
After five observations we switched to observing two fields simultaneously for an hour each, which meant we
could observe all four fields in 2 hours. From then until the end of the project we observed once per week.
The observations were carried out in blocks of 15\,min: 2\,min on the flux calibrator, followed by 11\,min on the target field.
The remaining 2\,$\times$\,1\,min were spent switching from the calibrator to the target and vice versa. By the end of the run,
four or eight snapshots (i.e. 44 or 88\,min on-source) had been obtained. We used 3C196 to calibrate the MD03 and MD05 data;
3C295 was the calibrator for MD06 and MD07.
The exact list of observing dates is reported in Table~\ref{tab:observing time}.

The total bandwidth per observation was 48\,MHz, covered by a total of 244
sub-bands, each with a bandwidth of 195.3\,kHz. We split this bandwidth in two to cover two target fields simultaneously.
We also used one sub-band per field to have a station beam in the middle of the two target fields in order to have them at
the same angular distance. In total we therefore used 121 sub-bands per field.
The sub-bands were grouped in six bands, which are neither equally spaced nor of equal width to avoid a priori known areas
of strong radio-frequency interference \citep[RFI, see][]{AOflagger_Offringa2010_bis, AOflagger_Offringa2012_bis}.
A description of these bands can be found in Table~\ref{tab:bands}.

\begin{table*}
\begin{center}
\begin{tabular}{|c|c|c|c|c|}
\hline
\hline
Run	& Target	& Date	& Total time on each field (min)		& Observation IDs	\\
\hline
1	& MD03-05		& 2013-03-02	& 88	& L99171 -- L99194	\\
2	& MD03-05		& 2013-03-30	& 88	& L111528 -- L111551	\\
3	& MD03-05		& 2013-04-13	& 88	& L119562 -- L119585	\\
4	& MD03-05		& 2013-04-27	& 88	& L126983 -- L127006	\\
5	& MD03-05		& 2013-05-11	& 88	& L133205 -- L133228	\\
\hline
6	& MD03-05-06-07	& 2013-06-08	& 44	& L144196 -- L144219	\\
7	& MD03-05-06-07	& 2013-06-16	& 44	& L146457 -- L146480	\\
8	& MD03-05-06-07	& 2013-06-30	& 44	& L151409 -- L151432	\\
9	& MD03-05-06-07	& 2013-07-12	& 44	& L151385 -- L151408	\\
10	& MD03-05-06-07	& 2013-07-20	& 44	& L151811 -- L151834	\\
11	& MD03-05-06-07	& 2013-07-25	& 44	& L151787 -- L151810	\\
12	& MD03-05-06-07	& 2013-08-10	& 44	& L151739 -- L151762	\\
13	& MD03-05-06-07	& 2013-08-17	& 44	& L151715 -- L151738	\\
14	& MD03-05-06-07	& 2013-08-24	& 44	& L169212 -- L169235	\\
15	& MD03-05-06-07	& 2013-08-30	& 44	& L172191 -- L172214	\\
\hline
\hline
\end{tabular}
\caption{Date, targets, time on target and observation IDs of all our observations.
Each snapshot is composed of three observation IDs: one for each field and one
for the calibrator observation before it.}
\label{tab:observing time}
\end{center}
\end{table*}

\begin{table}
\begin{center}
\begin{tabular}{|c|c|c|}
\hline
\hline
\multirow{2}{*}{Band number}	& Number of sub-bands & \multirow{2}{*}{Frequency range} \\
						&  per band	\\
\hline
0	& 20	& 115 -- 119 MHz \\
1	& 20	& 122 -- 126 MHz \\
2	& 20	& 126 -- 130 MHz \\
3	& 20	& 140 -- 144 MHz \\
4	& 20	& 154 -- 158 MHz \\
5	& 21	& 182 -- 188 MHz \\
\hline
\hline
\end{tabular}
\caption{Description of bands and bandwidth used in our observations.}
\label{tab:bands}
\end{center}
\end{table}

\subsection{Data Analysis}
\label{sec:analysis}

For all observations, data were recorded with a time and a frequency resolution of 2\,s and 3.05\,kHz (64~channels/sub-band),
respectively.
Pre-processing was carried out using standard methods. First, RFI was removed
using AOFlagger \citep[][]{AOflagger_Offringa2010_bis, AOflagger_Offringa2012_bis}.
Then, for practical reasons concerning data volume and the computing time required for calibration and imaging,
the data were averaged in time and frequency. After this step the data had time and frequency resolutions of 10\,s and
48.8\,kHz (4~channels/sub-band), respectively.
Time-averaging and bandwidth smearing have a minor effect on the field-of-view determination and calibration,
especially because we are analysing the inner 15.48\,deg$^2$ only \citep[see e.g.][]{Heald2014_MSSS}.

Calibration and imaging were carried out using standard practices \citep[][]{LOFAR_Heald2011,
LOFAR_vanHaarlem2013}.
The calibrator sub-bands were calibrated using a model of the source \citep[see][]{calibrators_Scaife2012}; the gain
amplitudes and phases were then transferred to the target field data.
After this step, we refined the calibration by performing phase-only calibration
on the target field using data from the LOFAR Global Sky Model \citep[see][for further details]{bartThesis_Scheers2011};
the basis for our model of the field was the 74\,MHz VLA Low-Frequency Sky Survey \citep[VLSS;][]{VLSS_Cohen2007},
with spectral index information being obtained by cross-correlating the relevant VLSS catalogue entries with the 1.4\,GHz
NRAO VLA Sky Survey \citep[NVSS;][]{NVSS_Condon1998} and the Westerbork Northern Sky Survey
\citep[WENSS;][]{WENSS_Rengelink1997}.

\begin{figure}
\begin{center}
\includegraphics[scale=0.3,viewport=5 5 792 737,clip]{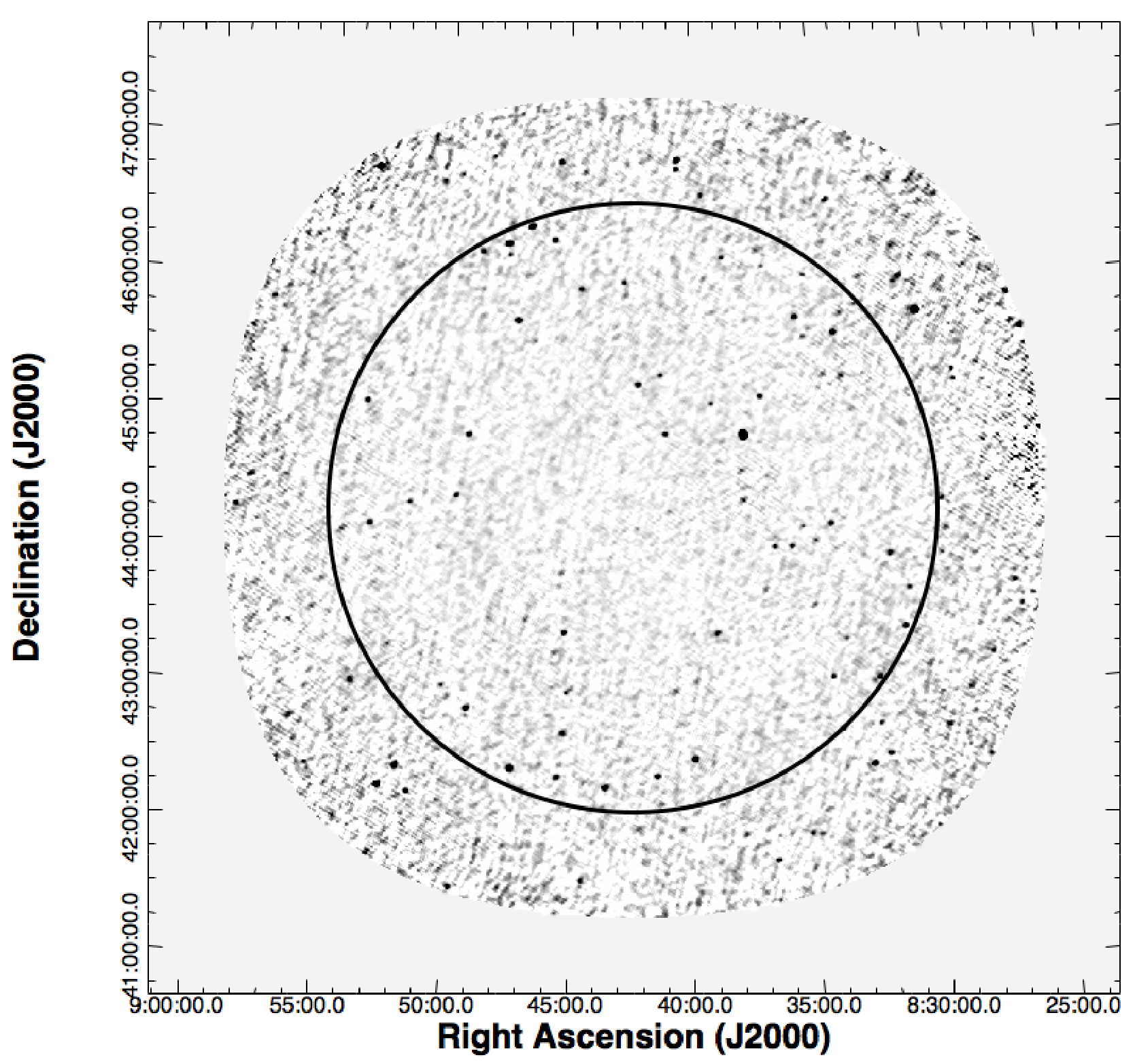}
\caption{Example of an image obtained after all the steps described in Section~\ref{sec:analysis} have been performed.
The black circle represents the area in which source extraction has been performed. This is an image of the MD03 field
with an average noise in the central part of 31\,mJy\,beam$^{-1}$ and an angular resolution of 1.8\,arcmin.}
\label{fig:avg_image}
\end{center}
\end{figure}

Primary-beam-corrected images were then made for each band using the {\sc awimager} \citep[][]{awimager_Tasse2013}.
We used a robust weighting parameter of 0 \citep[][]{imagingweight_Briggs1995}.
We used a maximum projected baseline length of 6000\,m; this cutoff gives the most reliable images for the relatively limited (u,v)
coverage and simple calibration strategy. The average noise level over the inner 15.48\,deg$^{2}$ of our images is
30\,mJy\,beam$^{-1}$, with an average angular resolution of 2\,arcmin.
An example of an image obtained after the processes described is shown in Figure~\ref{fig:avg_image}.

We decided not to apply direction-dependent effects as they are still computationally expensive at the time of writing.
The flux densities we obtain with LOFAR using this technique are comparable to those in other catalogues for a selection
of bright sources, for example the 7C survey at 151\,MHz \citep[][]{7C_Hales2007}.
We also note that \citet{Heald2014_MSSS} found a small flux scale offset (about 10\%) between LOFAR and 7C.

Comparing the sensitivity of our survey with others at different frequencies, we note that our survey is as sensitive
as the NVSS (average noise level of 0.45\,mJy\,beam$^{-1}$ at 1.4\,GHz) if the spectral index\footnote{We define the
spectral index $\alpha$ as: S($\nu$) $\propto$ $\nu^{\alpha}$.}  of the transients is $-$1.9, and more sensitive if it
is steeper. This further suggests that our survey may be particularly competitive at finding very steep
spectrum, coherent transients (also see Section~\ref{sec:intro}).

We are aware that not all of the transients are detectable with such a survey, both because of the timescale to which we are
sensitive and because of the flux sensitivity.
With this survey we are sensitive mostly to transients with time-scale between 10 minutes, corresponding to the duration of one
snapshot, and 5 months, corresponding to the duration of the observing campaign.
Transients shorter than one snapshot can still be detected if they are sufficiently bright such that the flux density
at the corresponding position, averaged over the snapshot, is still above the detection threshold. Another possibility
is an intrinsically short coherent burst that is highly affected by scattering, making it visible in the image plane at low radio frequencies.
One speculation about the newly discovered transient ILT\,J225347+862146 is, in fact, that it is a highly scattered or dispersed
FRB, which would then, however, require a fluence orders of magnitude greater than the ones found at GHz frequencies
\citep{Adam_Stewart2015}  and thus either a very broad luminosity distribution of such sources, or some with rather steep
spectra.
Transients with time-scales longer than the duration of the observing campaign will of course not be recognised as variable.
Therefore, the types of source that might be detectable by our survey range from transients such as GCRT~J1745$-$3009
\citep{GCRT1745_Hyman2005}, and ILT J225347+862146 \citep{Adam_Stewart2015}, to flare stars, X-ray binaries, and AGN
\citep[see][]{timescale_Tb_Pietka2015}.

\section{Results}
\label{sec:results}

\begin{figure}
\begin{center}
\includegraphics[scale=0.467,viewport=15 0 792 400,clip]{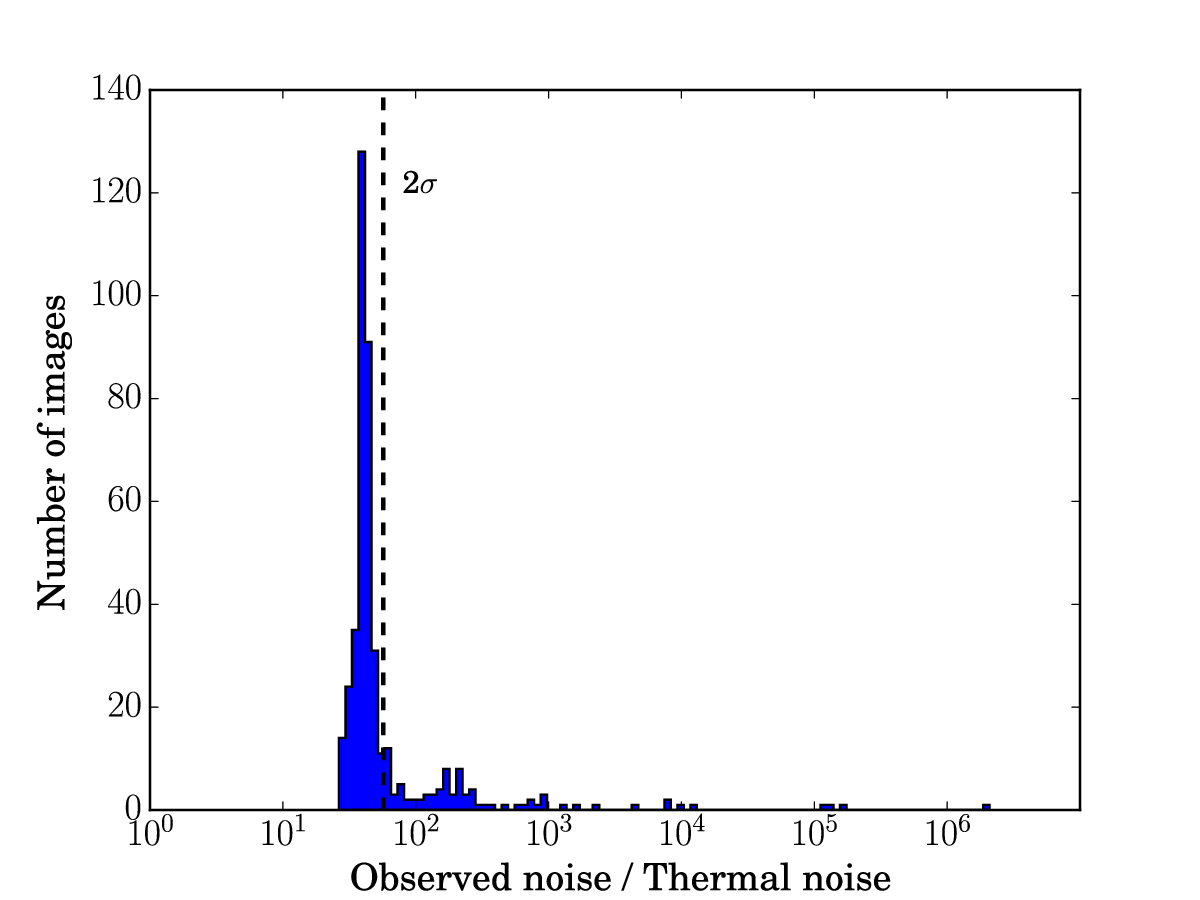}
\caption{Example of a plot from the quality control step for MD03. This plot shows a histogram of the ratio between
the measured noise in an image and the thermal noise. The histogram is then fitted with a Gaussian and a cut-off at
2\,$\sigma$ above the mean is made. All the images lying above 2\,$\sigma$ above the mean are discarded by the
pipeline and not considered for further analysis.}
\label{fig:QC_noise}
\end{center}
\end{figure}

A total of 1212 images from the four monitored fields at various observing times and frequencies have been
collected and processed by the Transients Pipeline \citep[TraP, ][]{Trap_Swinbank2015}.
A description of the main features of the pipeline is as follows. Firstly, a quality control step is used to eliminate bad images.
This check allows us to discard images with excessively high noise levels and excessively elongated beam shapes.
The criterion to reject images with a high noise level takes into account the ratio between the noise measured in the image
and the theoretical thermal noise, which is calculated from the integration time, the bandwidth and the antenna set.
We decided to use this ratio to compensate for the different observing conditions in different images (for example a different
number of stations could have been flagged out due to malfunctions or RFI).
To set the threshold for rejection, we created a histogram of the noise ratio of the images in the dataset for each field
separately. We fitted a Gaussian to the histogram and set a cut-off threshold at 2\,$\sigma$ above the mean. This means that
the threshold is different for different fields. An example of such a histogram is shown in Figure~\ref{fig:QC_noise} for the
case of MD03. For the beam ellipticity we calculated the ratio between the major and the minor axes of the beam for every
image in the dataset, and calculated the average and the RMS of the distribution for each field separately.
A cut-off at average+RMS was applied. The values used for the rejection are summarised in Table~\ref{tab:obs fields}.
Images with high noise or a highly elliptical beam were rejected since these are symptomatic of poor calibration
or RFI subtraction.

After the quality control process, 809 images remained. The MD05 field had a relatively high level of image rejection.
This was because at the start of the observation campaign the station beam was misaligned; as a result MD05 fell at the
edge of the beam, resulting in noisy images. This issue was solved after run 5.
A summary of the images can be found in Table~\ref{tab:obs fields}.

The remaining images have been searched for sources.
For each image a background RMS map was calculated over the entire image. Pixels with values above 8\,$\sigma$,
i.e. eight times the noise measured from the RMS map, are detected as seeds of sources and associated with neighbouring
pixels with values above 3\,$\sigma$ in islands to form a full source. A fit to these sources is then performed using elliptical
Gaussians. As mentioned in Section~\ref{sec:observations}, only sources in the central 15.48\,deg$^{2}$ were analysed
because outside this region the flux calibration was unreliable. A conservative threshold of 8\,$\sigma$ was
chosen to prevent spurious detections due to noise fluctuations. For an 8\,$\sigma$ detection threshold and purely
Gaussian noise, we expect fewer than 10$^{-7}$ false positive detections in the whole survey. Lower values for the
detection threshold have been tested but the number of spurious sources increased dramatically, indicating that the
noise is not purely Gaussian (as is known to be the case in radio images).

After source extraction, a database was populated with the measured properties and metadata of the
extracted sources. The source properties include position, peak and integrated flux, Gaussian fitting
parameters, and errors in all these quantities. The metadata include the time of observation, effective frequency and
beam properties \citep[][]{Trap_Swinbank2015}.
When a new source is extracted, the TraP establishes whether it should have been detected in previous images or not,
taking into account the different noise levels in different images.
In the case that it should have been detected but it was not, the source is labelled as a transient.

The TraP is also able to determine whether the flux of a source varies significantly during the survey.
To quantify the variability of sources we used two indicators: V$_{\nu}$ and $\eta_{\nu}$ \citep[][]{bartThesis_Scheers2011}.
The former indicates the relative magnitude of variability. It is expressed as
the ratio of the standard deviation (s$_{\nu}$) to the arithmetic mean ($\overline{I_{\nu}}$) of the flux over a sample of
$N$ measurements of a source:

\begin{equation}
V_{\nu}\,=\,\frac{s_{\nu}}{\overline{I_{\nu}}} = \frac{1}{\overline{I_{\nu}}} \sqrt{\frac{N}{N-1} \left(\overline{I_{\nu}^{2}} -
\overline{I_{\nu}}^{2}\right)}\ .
\label{eq:Vnu}
\end{equation}

The second indicator, which expresses the significance of the flux variability, is based on reduced $\chi^{2}$ statistics.
It indicates how well a constant value fits the light curve of a source, and thus how variable the light curve is:

\begin{equation}
\eta_{\nu}\,=\,\frac{N}{N-1} \left(\overline{w I_{\nu}^{2}} - \frac{\overline{wI_{\nu}}^{2}}{\overline{w}}\right) ,
\label{eq:Etanu}
\end{equation}

\noindent
where $w$ are weights inversely proportional to the errors in the flux measurements $\left(w \propto 1/\sigma_{I_{\nu}}^{2}\right)$.

If a source has an outlier in one of its flux measurements, then V$_{\nu}$ will be large. However, the significance of $\eta_{\nu}$
depends on the errors of the flux measurements: when $\sigma_{I_{\nu}}$ is large, the significance will generally be weaker,
whereas small changes in the flux density can be highly significant if $\sigma_{I_{\nu}}$ is small.
Following Rowlinson et al. (in prep.),
we used a threshold to select transient candidates: a histogram of
each indicator is built and fit with a Gaussian in logarithmic space. A cut at 3\,$\sigma$ above the mean is used to select
variable sources. In this work we are focusing on transient sources: sources which are detected in only one (or a few)
snapshots. After this analysis no credible transient candidates were found in our survey.

\begin{figure}
\begin{center}
\includegraphics[scale=0.467,viewport=15 0 792 400,clip]{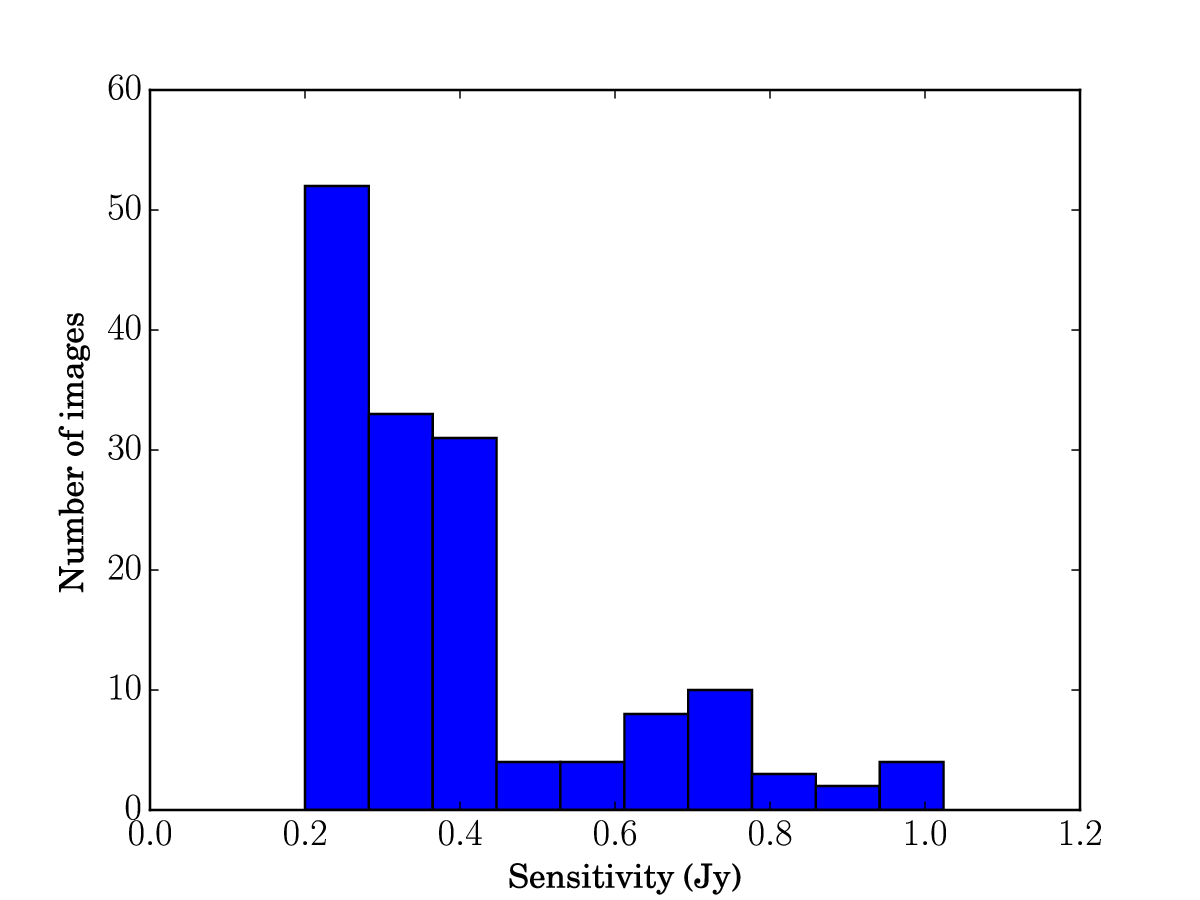}
\caption{
Histogram of the distribution of images as a function of their sensitivity. We note that most of the images have a sensitivity
below 0.4\,Jy, but there is a tail of images that are more noisy.
}
\label{fig:histogram}
\end{center}
\end{figure}

\section{Methods to determine the transient surface density limits}
\label{sec:methods}

As we detected no radio transients with this survey, we will use the total area that was sampled to constrain the transient
surface density, i.e. the number of transient sources that are brighter than a certain flux per
square degree at any instant.

Establishing the transient surface density of requires quantifying the total field of view of the survey.
Each snapshot of each field covers a total field of view of 27.92\,deg$^2$ in a circular region around the phase centre.
However, only the inner 15.48\,deg$^2$ have been searched for transients, as the outer part of the field of view is much
noisier and the flux calibration less reliable. 
Each comparison between independent consecutive images of the same field of view increases the survey area by an increment
equal to the image area. Thus, two 15.48\,deg$^2$ images of the same field provides a single comparison and a total
survey area of 15.48\,deg$^2$; three images provide two (consecutive) comparisons, and a survey of 30.96\,deg$^2$,
and so on.
We define independent images as those which share no overlap with others in the time domain. Thus, multiple images
of the same field at the same time at different frequencies are not independent. Therefore, we count as one independent
image each snapshot for which at least one band passed quality control.
Using these metrics, we have 151 independent images, for a total survey area of
(151\,--\,4)\,$\times$\,15.48\,deg$^2$\,=\,2275\,deg$^2$.
In Figure 3 we show a histogram of the sensitivities of the remaining images. One can see that most of our observations
have sensitivities below 0.4\,Jy, but there is a tail of images that are less sensitive.

We used three methods to calculate the transient surface density upper limit.
It is transient surface density and not transient rate because the quantity we are determining is the number of transient
sources per unit area (deg$^{2}$) that we find in an image. This quantity is not per unit time, and therefore it is not a rate.

We first determine the transient surface density following the traditional method that has been used by many authors
before us, which gives the upper limit at one specific flux. We expand upon this method by calculating upper limits at
different fluxes in two ways: one model-independent method, and another one in which we assume that the transient
source population follows a power-law distribution as a function of flux. The first two methods
(Sections~\ref{sec:old_method} and \ref{sec:method1}) and the third method (Section~\ref{sec:method2}) are independent
in the sense that the latter can take into account the variable sensitivity in different images, while this is not possible in
the first two methods. 
n Figure~\ref{fig:methods} we will populate the transient surface density vs flux density plane with the upper limits we
will derive in Sections~\ref{sec:old_method}, \ref{sec:method1}, and \ref{sec:method2}.

\subsection{``Traditional'' transient surface density}
\label{sec:old_method}
To calculate the 95\% confidence level upper limit to the transient surface density from our survey we
assume a Poisson distribution:

\begin{equation}
P(n)\,=\,\frac{\lambda^{n}}{n!} e^{-\lambda} .
\label{eq:poisson}
\end{equation}

Rewriting the Poisson variable $\lambda$ as the product of the surface density and the total
number of square degrees sampled in this survey ($\lambda$\,=\,$\rho$\,$\Omega_{\mathrm{tot}}$), and assuming no transients
were detected ($n$\,=\,0), we can rewrite the equation as:

\begin{equation}
P(0)\,=\,e^{-\rho \: \Omega_{\mathrm{tot}}} ,
\label{eq:poisson no transient}
\end{equation}

\noindent
where $\rho$ is the transient surface density and $\Omega_{\mathrm{tot}}$ is the total amount of solid
angle scanned during our survey.
The 95\% confidence level is defined as $P(0)$\,=\,0.05\footnote{Incidentally, since -ln(0.05)=3.00, if one finds zero
sources in any survey one should quote an upper limit to the expectation value of the number of detections of 3 to
be even moderately conservative, not 1 (as is often done).}.
Since $\Omega_{\mathrm{tot}}$\,=\,2275\,deg$^2$ we get a limit of $\rho$\,$<$\,1.28\,$\cdot$\,10$^{-3}$\,deg$^{-2}$.
The flux limit at which this limit is calculated is determined by the detection limit on the noisiest image in our sample.
As mentioned in Section~\ref{sec:results}, we used a detection threshold of 8\,$\sigma$ which translates to a flux limit
of about 1\,Jy. This is indicated with a star in Figure~\ref{fig:methods}.
This is the analysis other surveys in the literature performed to calculate transient surface density
\citep[see for example][]{VLAarchival_Bell2011, SN&trans_Alexander2014}.
Restricting our analysis to the noisiest image is not the best approach because, as shown in Figure~\ref{fig:histogram},
most of our observations are much better than the one that is setting our flux limit using this approach.

\subsection{Beyond the traditional method}
\label{sec:method1}

We can expand this method to calculate upper limits of the transient surface density at different fluxes.
Instead of using all the observations, we can also eliminate the noisiest image from our dataset, gaining a
better sensitivity, but over a reduced survey area, followed by repeating the
same calculation. We can iterate this method using fewer and fewer images with better and better overall sensitivity.
The limits set with this method are illustrated in Figure~\ref{fig:methods} with grey dots.
This method is free of a priori assumptions on the source flux distribution, just as the traditional method, but has the
advantage over the former that it uses all the information in the images to produce an entire curve of limits below
which any population of transients must lie. As we shall see in the next section, which part of the curve is most
constraining on any population depends on its flux distribution.
Using this approach we are able to explore a larger portion of the flux axis, the portion allowed by the sensitivity of our
observations. This method is still not optimal because at every iteration we are treating all of the images as having the
sensitivity of the noisiest in the remaining dataset, whereas many of them are much more sensitive.

\subsection{Transient surface density versus flux distribution}
\label{sec:method2}
We can get more stringent upper limits if we assume that the number density of transient sources has a power-law
distribution of flux densities:

\begin{equation}
N(S > \hat{S})\,=\,N_{*} \left(\frac{\hat{S}}{S_{*}}\right)^{-\gamma} ,
\label{eq:flux distribution}
\end{equation}

\noindent
where $N_{*}$ is the normalisation and $S_{*}$ is an arbitrary value of the flux at which the normalisation is given.
It is sensible to use a value of $S_{*}$ within the range of fluxes we measured in our survey.
This assumption is justified by the fact that sources that might be the progenitors of radio transients generally have
a power-law distribution in flux, for example gamma-ray burst afterglows, active galactic nuclei, etc. and therefore
the cumulative flux distribution of transient sources will reasonably follow a power law as well.
Also, the range of sensitivities explored in a given survey usually spans only a factor few, so any broad flux distribution
can be reasonably approximated by a power law.
In this case we can estimate how many transient sources we should have seen in an image with noise $\sigma_{i}$
and a signal-to-noise threshold \emph{D$_{i}$}. This number is just the number density of sources brighter than
$D_{i}\,\sigma_{i}$ multiplied by the field of view of the image ($\Omega_{i}$).

\begin{equation}
n_{i}\,(S > D\,\sigma_{i})\,=\,N_{*}  \left(\frac{D_{i} \,\sigma_{i}}{S_{*}}\right)^{-\gamma}  \Omega_{i} \ .
\label{eq:sources per image}
\end{equation}

\noindent
Adding up the number of transient sources that we should have seen in our whole dataset, we end up with the expression:

\begin{equation}
n_{\textrm{tot}}\,=\,\Sigma_{i} n_{i} = N_{*} \: \sum_{i} \left(\frac{D_{i}\,\sigma_{i}}{S_{*}}\right)^{-\gamma} \Omega_{i} \ .
\label{eq:total number}
\end{equation}

In our case, the signal-to-noise threshold and the field of view are the same for all our images and can be taken out of
the summation  (from now on they will be identified by \emph{D} and $\Omega$). 
We can now calculate the upper limit of the transient surface density at the 95\% confidence
level, assuming a Poisson distribution as in Equation~\ref{eq:poisson}, and write:

\begin{equation}
\textrm{exp}(-n_{\textrm{tot}})\,=\,\textrm{exp}\left(-N_{*} \: \left(\frac{D}{S_{*}}\right)^{-\gamma} \: \Omega \: \sum_{i}\sigma_{i}
^{-\gamma}\right) = 0.05 \ .
\label{eq:upper limit}
\end{equation}

Solving for the normalisation $N_{*}$ of the transient source flux distribution we get:

\begin{equation}
N_{*}\,<\,-\frac{\textrm{ln}(0.05)}{\Omega} \: \left(\frac{S_{*}}{D}\right)^{-\gamma} \: \frac{1}{\sum_{i}\sigma_{i}^{-\gamma}} \ .
\label{eq:k upper limit}
\end{equation}

This means that we obtain an upper limit to the number of transient sources as a function of the signal to noise
threshold of the survey for any given value of the exponent of the flux distribution ($\gamma$).
These functions are displayed as lines in Figure~\ref{fig:methods} for values of $\gamma$ from 0 to 2.5.
The choice of the flux at which to give the normalisation ($S_{*}$) is completely arbitrary.
In Table~\ref{tab:new_rate} we quote the value of the upper limits we calculate for different values of $\gamma$
at $S_{*}$ = 0.5\,Jy because it falls in the range of sensitivities we sampled. These values are represented with
dots in Figure~\ref{fig:methods}.
Note that with $\gamma$ = 0 we recover the result from the method described in Section \ref{sec:old_method}.

This method allows us to put more stringent upper limits on the transient surface density as we are using all the information
in our data. The price we pay is that we need to make an assumption on the transients flux distribution, which means
that our results are valid within the boundaries where our assumption holds. We do not think this assumption is severe,
since the power-law approximation only needs to hold over a factor of three in flux density (i.e. the range of the various
measurements). One should therefore not use the power-law approximation to extrapolate the limits we derive too much
outside the flux interval we probed.

\begin{table}
\begin{center}
\begin{tabular}{|c|c|c|}
\hline
\hline
\multirow{2}{*}{S$_{*}$ (Jy)} & \multirow{2}{*}{$\gamma$} 	& transient surface density \\
						&						&upper limit (deg$^{-2}$)	\\
\hline
0.5	&	2.5	&	3.6$\,\cdot\,10^{-4}$	\\
0.5	&	2.0	&	4.9 $\,\cdot\,10^{-4}$	\\
0.5	&	1.5	&	6.5 $\,\cdot\,10^{-4}$	\\
0.5	&	1.0	&	8.5 $\,\cdot\,10^{-4}$	\\
0.5	&	0.5	&	1.06 $\,\cdot\,10^{-3}$	\\
0.5	&	0.0	&	1.28 $\,\cdot\,10^{-3}$	\\
\hline
\hline
\end{tabular}
\caption{Upper limits on the transient surface density from our survey for different values of the exponent of the
assumed flux distribution of transient sources. See Equation~\ref{eq:flux distribution}.}
\label{tab:new_rate}
\end{center}
\end{table}

\subsection{Analysing the methods}
\label{sec:methods analysis}
All the values quoted so far take into account every snapshot as an independent observation and therefore are valid for
transients with characteristic time-scale of 15 minutes. This is the dynamical time-scale near a black hole with a horizon
radius of 2\,AU, or mass $10^8\,M_\odot$, but since black-hole sources show variability on a wide range of time scales
this gives no precise constraints on the sources.
To derive proper limits on the surface density and rate at longer time-scales, consecutive snapshots must be merged as
will be explained in Section~\ref{sec:timescale}.

\begin{figure*}
\begin{center}
\includegraphics[scale=0.60,viewport=0 0 530 403,clip]{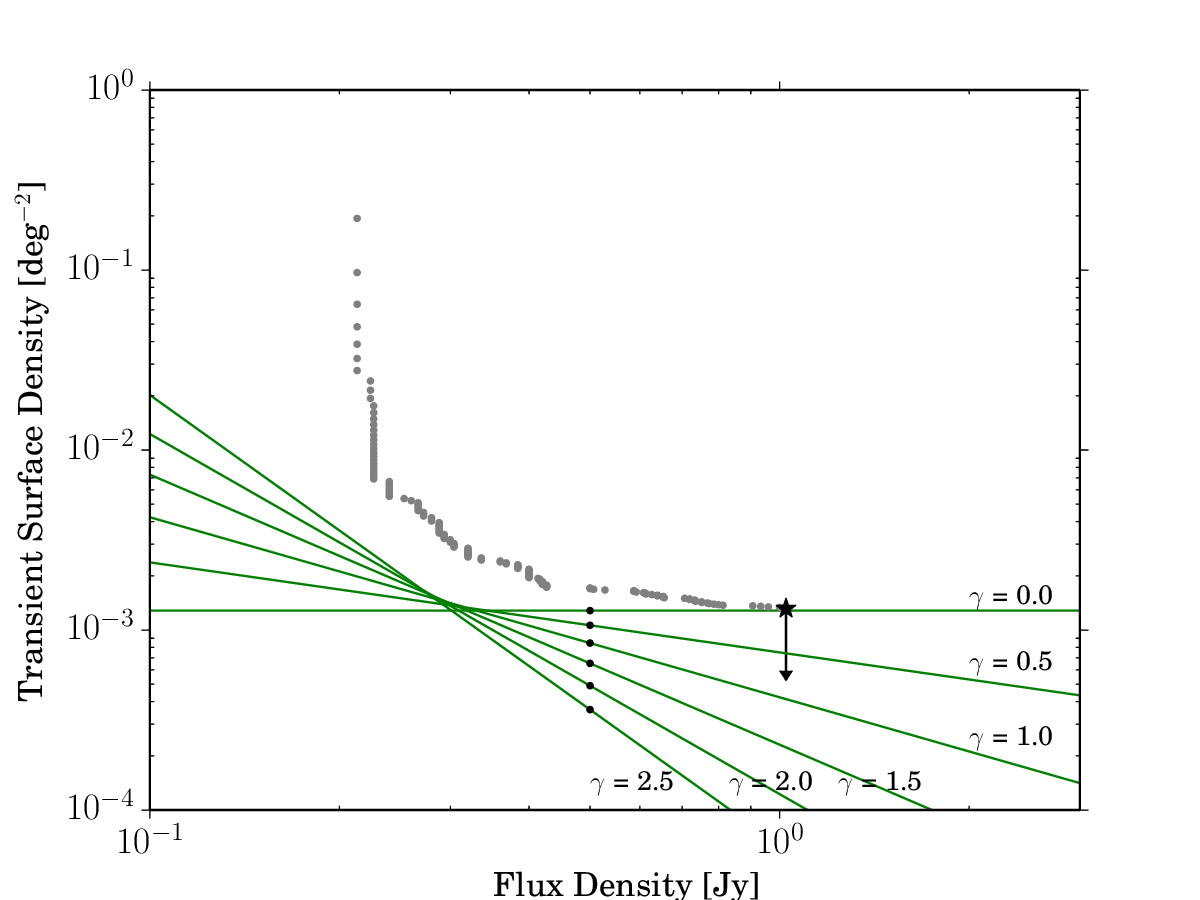}
\caption{Transient surface density $\left(\mbox{deg}^{-2}\right)$ against flux density (Jy) derived with different methods.
The star represents the result of the traditional method described in Section~\ref{sec:old_method}. The grey dots represent
the results obtained with the model independent method described in Section~\ref{sec:method1}. 
The lines represent the upper limits obtained with the method described in Section~\ref{sec:method2} which assumes
transient sources follow a power-law distribution in flux
(see Equation~\ref{eq:flux distribution}).
The normalisation of the lines is determined by the data; the flux at which to give the normalisation (S$_{*}$) is arbitrary.
The dots represent the values of the normalisation at S$_{*}$\,=\,0.5\,Jy, as reported in Table~\ref{tab:new_rate}.
The choice of a different value of S$_{*}$ would only move the dots left or right along the lines.
}
\label{fig:methods}
\end{center}
\end{figure*}

Finally, we compare the three methods discussed so far with each other. In Figure~\ref{fig:methods}, we show the naive,
traditional estimate of using the same flux limit for each image by a star; this limit is the poorest one among the
reasonable images, i.e., after removing the outliers from the sensitivity distribution as described in Section~\ref{sec:results}.
To this we add, with grey dots, the results of our iterative model-independent lowering of the flux limit by successive removal
of the highest-noise remaining image. The resulting curve has a fairly characteristic shape: at the highest fluxes it is fairly
flat, because we retained images up to $2\sigma$ above the mean RMS noise, and so initially we lower the flux limit significantly
while losing only few images (i.e., little survey area). At lower fluxes the opposite happens: the curve is very steep, because
we run up against the minimum possible noise and there are few images left, so we suffer a high loss of fractional survey area
with little gain in flux limit. The optimum combination of flux limit and transient surface density limit is somewhere in the middle.
We can gain more insight into that optimum by adding the results of our analysis using the assumption of a power-law distribution
of source fluxes. These results are plotted in Figure~\ref{fig:methods} as the lines labelled by their values of $\gamma$
and with the fiducial point at $S_{*}$\,=\,0.5\,Jy marked as a dot. Again, we note that the value of $S_{*}$ is an arbitrary choice;
any other point on each curve may serve as well. We see that indeed each curve lies completely below the grey
dots, confirming that making a model assumption about the population leads to a stricter limit (with equality at the unphysical case
$\gamma$\,=\,0).
It is easy to see why the curve lies clearly below the grey dots everywhere: each of the grey dots is a non-optimum
summary of the data: at the high-flux end, it uses a too conservative flux limit for most of the survey, and at the low-flux end it uses
only a small fraction of the survey area. One can also see why the ordering of the curves is as it is: at the high end, the shallow
source-count populations come closest to the grey dots, because most sources in such a population are bright, so much less
damage is done to the quality of the limit by setting a high threshold for all images. At the low end, the opposite is true, because for
very steep source counts most of the constraint is in the few best images, and discarding the less sensitive ones hurts little. Because
of this ordering, any two curves must intersect each other somewhere within the survey flux range; somewhat surprisingly,
they all intersect at about the same flux, and this is also the flux at which they come closest to the grey dots, i.e., where the previous
method has its optimum compromise between sensitivity per image and area covered. At this flux value, $S_{\rm best}$\,=\,0.3\,Jy,
we will quote our overall, optimal transient surface density limits (this flux is also the mode of the single-image sensitivity distribution):
from method 2, the grey dots, we find a completely model independent surface density limit of $5\,\cdot\,10^{-3}$\,deg$^{-2}$.
From the fits with a power-law source count model, we find a surface density limit of
$1.3\,\cdot\,10^{-3}$\,deg$^{-2}$. This limit is, in the end, also virtually model-independent, since it is the same for all values of the
source count slope. This value we therefore quote as our overall, and robust, transient surface density limit.
Note that it is a factor 3 deeper in flux than can be obtained from previously used, simpler methods in the literature, and a factor 4 smaller
in transient surface density than can be obtained from the model independent methods.

\subsection{Transient surface density at different time-scales}
\label{sec:timescale}

It should be noted that the time axis is not included in the analysis performed so far.
At lower frequencies the time-scales of flux variation are usually longer for incoherent sources
\citep[][]{timescale_frequency_VanDerLaan1966}.
In our survey we are sensitive to time-scales ranging between 15 minutes (the time difference between two consecutive
snapshots) and about 5 months (the difference between the first and last observation).
To determine how our upper limit on the transient surface density is changing as a function of the time-scale, we
computed how many pairs of observations we have at specific time separations.
We explored time-scales of 15, 30, 45, 60, 75, 90, and 105 minutes (corresponding to 1, 2, 3, 4, 5, 6, and 7 snapshot
differences, within the same observing day), one week, two weeks, one month, and two months.

To do this, we binned together images with a time difference which was lower than the time-scale of interest.
This means that all snapshots from the same observation collapse into one measurement for time-scales longer than a day,
two observations within a week are merged for time-scales longer longer than a week, and so on. 
Pairs of the remaining observations separated by a time difference equal to the time-scale of interest are created.
We did not combine the images to create a deeper map, but considered the flux limit as the flux of the worst image in the dataset.
Their number depends on the time-scale, N$_{\textrm{pairs}}$(T), and is converted into a surveyed area, and then
into a number of sources, multiplying it by the field of view of each snapshot and by a transient surface density.
We then calculate an upper limit on the transient surface density using Equation~\ref{eq:poisson no transient} resulting
in the expression

\begin{equation}
\rho(T)\,<\,-\frac{\textrm{ln}(0.05)}{N_{\textrm{pairs}}(T)\,\cdot\,\Omega} \ .
\label{eq:timescales}
\end{equation}

\begin{figure*}
\begin{center}
\includegraphics[scale=0.45,viewport=0 0 522 403,clip]{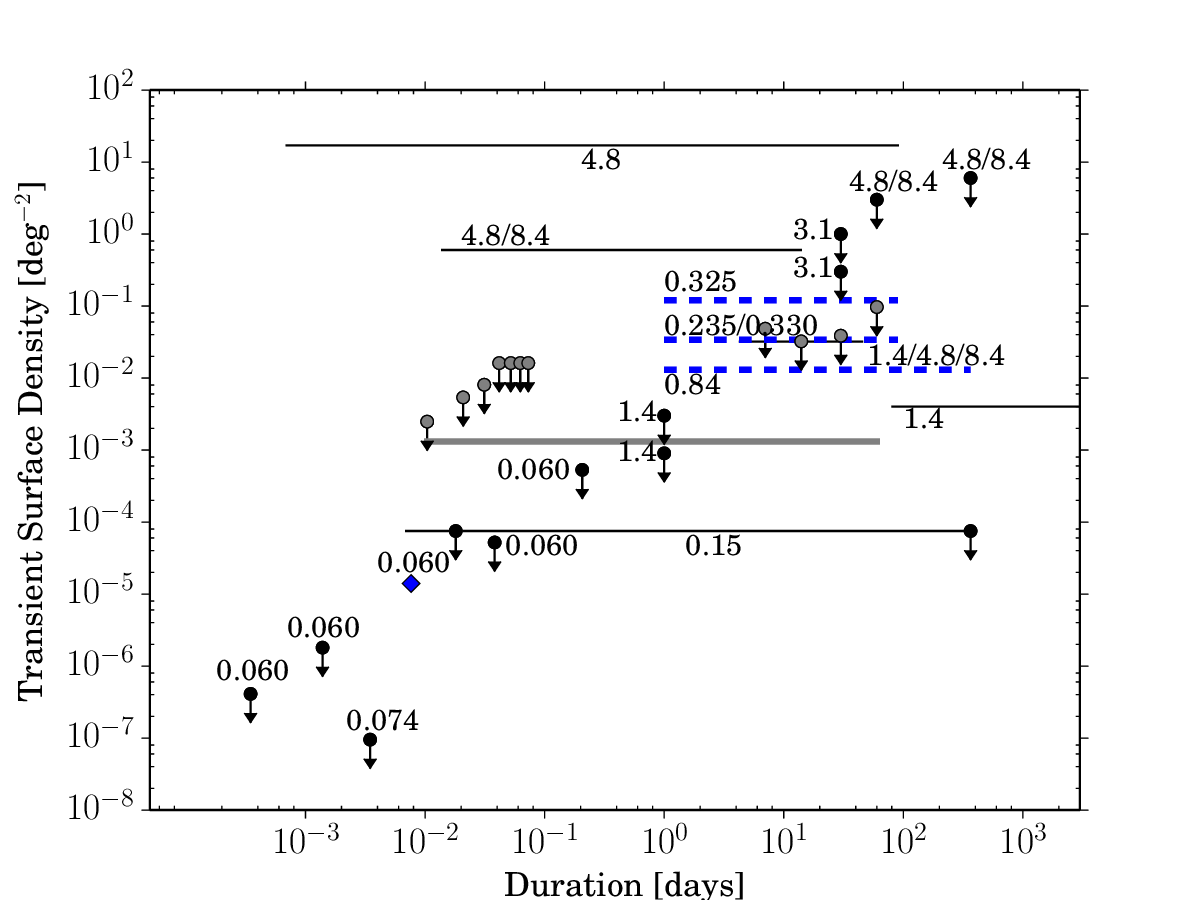}
\includegraphics[scale=0.45,viewport=0 0 522 403,clip]{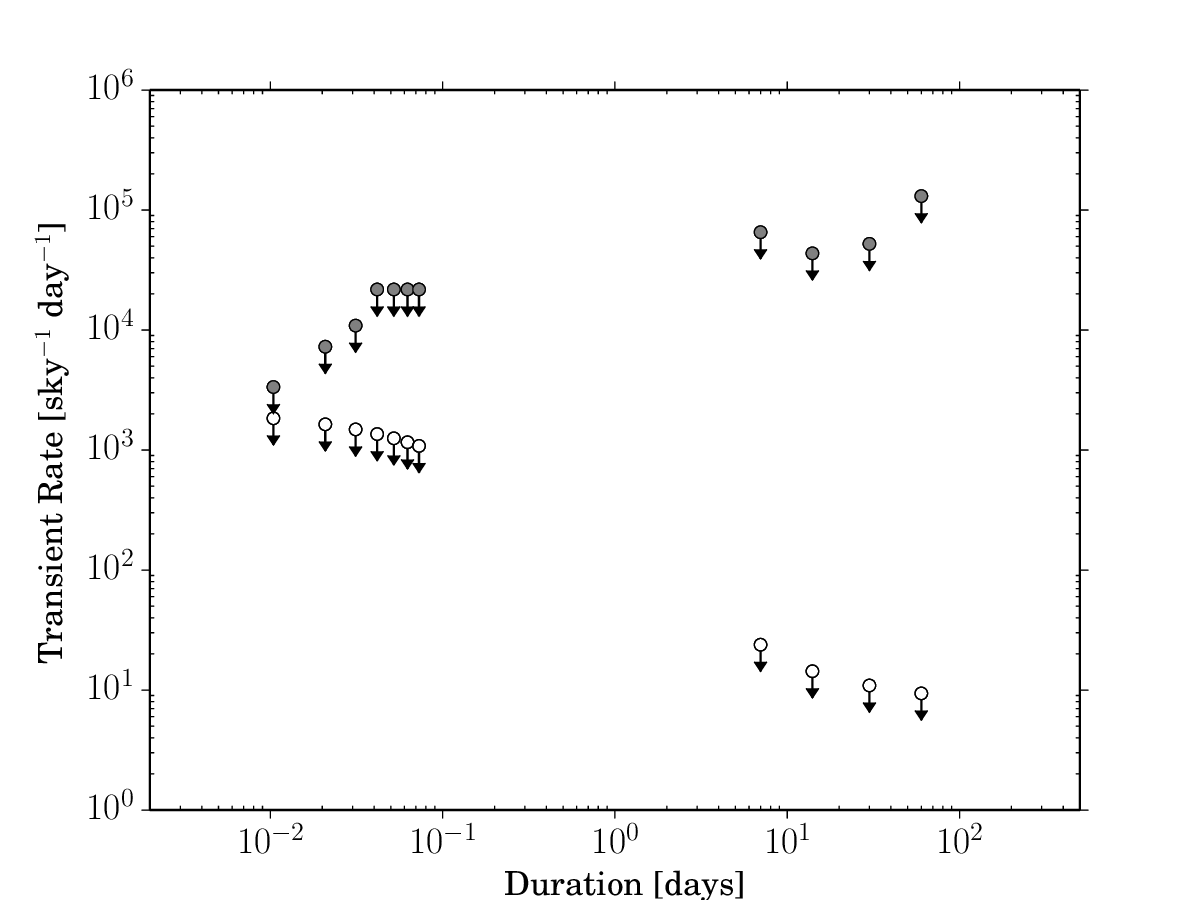}
\caption{Transient surface density $\left(\mbox{deg}^{-2}\right)$ against time-scale based on detections of transients
(dashed lines and diamond) and upper limits based on non-detections (circles and lines). The numbers indicates
the frequency in GHz at which each survey was conducted. The surveys displayed here are listed in
Table~\ref{tab:comparison}.
The grey circles indicate the upper limits derived from this work while the grey line represents the result from this
work as applied to the other surveys. It is clear that the grey line is only a very rough
(and too optimistic) approximation for the more correct values indicated by the grey
circles, underlining the importance of our more careful analysis.
In the right panel we plot the transient rate we can calculate with (open symbols) and without (filled symbols) applying
a correction to take into account the fact that transients could have fallen into gaps between observations as
described in Equation~\ref{eq:timescales_newmethod}.}
\label{fig:compare_timescales}
\end{center}
\end{figure*}

This gives us the number of transients for a given time-scale falling within one of our snapshots. 
Although this not entirely correct, this is what other studies in the literature have reported, and therefore we use this to
compare our results, which are plotted in the left panel of Figure~\ref{fig:compare_timescales}.
The horizontal thick grey line in this figure indicates the level at which we calculated the upper limit for the transient surface
density throughout the whole range of time-scales we probed, using the same method as in other studies.
The grey circles indicate the upper limits at different time-scales, showing that the upper limit we can set changes
dramatically at different time-scales. 
The left panel of Figure~\ref{fig:compare_timescales} shows that our survey is most sensitive to transients with a time-scale
of 15 minutes, at a level close to the horizontal grey line, while we are less sensitive to transients on the order of months.
This difference spans almost two orders of magnitude in surface density, and it is evident how a constant surface density
limit is a very rough approximation of the data at almost any time-scale.
It is also clear from this figure that we can not set any limits for transients with time-scales on the order of several hours
to a few days as we do not have any pairs of observations covering those time-scales. 
In the same figure we compare our results with those of the surveys reported in Table~\ref{tab:comparison} that have
reported a time-scale (for further details about these surveys see Section~\ref{sec:discussion}).
In order to plot the data from those surveys we use a constant line plotted across the whole range of time-scales at
their reported transient surface density limit.
The left panel of Figure~\ref{fig:compare_timescales} shows how our survey is competitive with respect to others
at different, often much higher frequencies.

We were also able to calculate an upper limit on the transient rate as a function of the duration of the transients.
To do so we rewrote the Poisson variable $\lambda$ as the product of the transient rate, the surveyed area, and
the total observing time ($\lambda$\,=\,$\hat{\rho}$\,$\Omega_{tot}$\,t$_{obs}$).
The total surveyed area can now be express as: $\Omega_{tot}$(T)\,=\,N$_{fields}$\,$\cdot$\,$\Omega$.
The observing time can be expressed as a function of the duration of the transients as only statistically independent
pairs of observations give information on the transient rate. This quantity can therefore be expressed as:
t$_{obs}$\,=\,N$_{pairs}$(T)\,$\cdot$\,t$_{snap}$, where t$_{snap}$ is the duration of each observation.
We then calculate an upper limit on the transient rate as a function of the duration of the transients using
Equation~\ref{eq:poisson no transient} resulting in the expression:

\begin{equation}
\hat{\rho}_{obs}(T)\,<\,-\frac{\textrm{ln}(0.05)}{N_{\textrm{fields}}\,\cdot\,\Omega\,\cdot\,N_{pairs}(T)\,\cdot\,t_{snap}} \ .
\label{eq:rate_timescale}
\end{equation}

\noindent Results from this calculation are shown in the right panel of Figure~\ref{fig:compare_timescales} as filled circles.

We can also calculate the transient rate, dividing the total number of transients between the beginning and the end of
our survey by the amount of square degrees we surveyed and the total amount of time we were sensitive to a transient
with duration T.
The amount of area we surveyed is the the sum of the fields of view on each field
($\Omega_{tot}$\,=\,N$_{\textrm{fields}}$\,$\cdot$\,$\Omega$).
The total amount of time we were sensitive to a transient as a function of its duration 
we approximate as the total survey time plus one duration, meaning that the transient could have started
one duration before the first snapshot and still be detectable in it
(t$_{tot}$(T)\,=\,t$_{survey}$\,+\,T).
This is valid in the limit of very bright sources. A more detailed calculation should take into account the flux reduction
due to a transient falling only partially into an observation.
To get an estimate of the total number of transients between the beginning and the end of
our survey we have to correct for the probability of a source of duration T to fall completely
in gaps between two observations (P$_{\textrm{gaps}}$). This probability is equal to the ratio between the amount of time when
a transient can start without being detectable in any snapshot and the total duration of the survey.
A transient of duration T is detectable in a snapshot starting at $t_{\textrm{start}}$ and finishing
at $t_{\textrm{end}}$ if it starts between $t_{\textrm{start}}$\,--\,T and $t_{\textrm{end}}$. This implies that in a gap of length
$t_{\textrm{gap}}$ between two
consecutive observations the amount of time when a transient of duration T can start and not fall in any of the two snapshots
is equal to max[($t_{\textrm{gap}}$\,--\,T), 0]. Summing this on all the gaps and dividing by the total survey time we obtain the
probability we were looking for:

\begin{equation}
P_{\textrm{gaps}}(T)\,=\,\frac{\sum_{i}\,\textrm{max}\left[(t_{\textrm{gap,\,i}} - T),\:0\right]}{t_{\textrm{survey}}} \ ,
\end{equation}

Now we can calculate the total number of transients between the beginning and the end of the survey as:

\begin{equation}
n_{\textrm{obs}}(T)\,=\,n_{\textrm{tot}}(T)\,\cdot\,(1\,-\,P_{\textrm{gaps}}(T)) \ ,
\end{equation}

\noindent where n$_{\textrm{obs}}$ is the number of transients detectable in at least one snapshot and n$_{tot}$ is the
total number of transients appearing between the beginning and the end of the survey.
These two parameters can be expressed as:

\begin{align*}
n_{\textrm{obs}}&=\,\hat{\rho}_{\textrm{obs}}\,\cdot\,N_{\textrm{fields}}\,\cdot\,\Omega\,\cdot\,N_{pairs}(T)\,\cdot\,t_{snap} \\
n_{\textrm{tot}}&=\,\hat{\rho}_{\textrm{tot}}\,\cdot\,N_{\textrm{fields}}\,\cdot\,\Omega\,\cdot\,(t_{survey}\,+\,T) \\
\end{align*}

\noindent and therefore:

\begin{equation}
\hat{\rho}_{\textrm{tot}}(T)\,=\,\hat{\rho}_{\textrm{obs}}(T)\frac{N_{pairs}(T)\,\cdot\,t_{snap}}{t_{survey}\,+\,T}\frac{1}{1\,-\,P_{\textrm{gaps}}(T)} \ ,
\end{equation}

\noindent where $\hat{\rho}_{\textrm{obs}}$ is given in Equation~{\ref{eq:rate_timescale}} and can be used to derive:

\begin{equation}
\hat{\rho}_{\textrm{tot}}(T)\,<\,-\frac{\textrm{ln}(0.05)}{N_{\textrm{fields}}\,\cdot\,\Omega\,\cdot\,(t_{survey}\,+\,T)}\frac{1}{1\,-\,P_{\textrm{gaps}}(T)} \ .
\label{eq:timescales_newmethod}
\end{equation}

This equation is simplified in three different regimes.
If T\,$\ll$\,t$_{\textrm{gap, i}}$ $\forall$ i:

\begin{align*}
&t_{\textrm{gap,\,i}} - T \approx t_{\textrm{gap,\,i}} \\
&\sum_{i}\,\textrm{max}\left[(t_{\textrm{gap,\,i}} - T),\: 0\right] \approx t_{\textrm{no\_obs}} \ , \numberthis
\end{align*}

\noindent where t$_{\textrm{no\_obs}}$ is the total amount of time between the beginning and the end of the survey when we
were not observing. \newline
If T\,$>$\,max(t$_{\textrm{gap}}$):

\begin{equation}
\textrm{max}\left[(t_{\textrm{gap,\,i}} - T),\:0\right]\,=\,0 \ \forall i \ ,
\end{equation}

\noindent therefore the probability of a transient to fall in gaps is null and no correction is needed. \newline
If 0\,$\le$\,t$_{\textrm{gap,\,i}}$ -- T $\forall$ i:

\begin{equation}
\sum_{i}\,\textrm{max}\left[(t_{\textrm{gap, i}} - T),\:0\right] = \frac{t_{\textrm{no\_obs}}\,-\,N_{\textrm{pairs}}(T)\cdot T}{t_{\textrm{survey}}} \ .
\end{equation}

We are aware that this correction is an approximation and that a Monte-Carlo analysis is required to fully address
this problem. We are exploring this in a follow-up study (Carbone et al., in prep).
The results of this correction are reported in the right panel of Figure~\ref{fig:compare_timescales} as empty circles.
There we compare the results with and without applying this correction.
We find that our survey is most sensitive to transients with time-scale of the order of months.
It is still clear how the upper limit we can set changes dramatically with the time-scale of the transients.
It is also clear that we can set more stringent upper limits for transients with longer duration due to the fact that
for longer durations, even if our sampling is not constant we are not losing any information by not observing:
it is not possible to miss a transient whose duration is longer than the longest gap between consecutive observations. 
For short transients this correction does not change the limit that can be set on the transient rate.
Transient rate limits are given at specific flux sensitivities and this information is not included in our figure.
We are aware that the relationship between flux sensitivity, transient rate and transient time-scale is more
complex than we can show on one single plot. This will also be taken into account in a follow-up study
(Carbone et al., in prep).

\section{Comparison with other surveys and other methods}
\label{sec:discussion}

A number of radio transient searches have been published in recent years. We summarise them here, and give their
main results in Table~\ref{tab:comparison} for comparison with our results.

\citet{ATATS_Croft2010} published results from the ATA Twenty Centimetre Survey (ATATS) at 1.4\,GHz, and
subsequently the Pi GHz Sky Survey (PiGSS) surveyed the sky with ATA at 3.1\,GHz \citep[][]{PiGSS_Bower2010}.
No transients were detected and an upper limit on the transient surface density was reported.

\begin{figure*}
\begin{center}
\includegraphics[scale=0.45,viewport=0 0 522 396,clip]{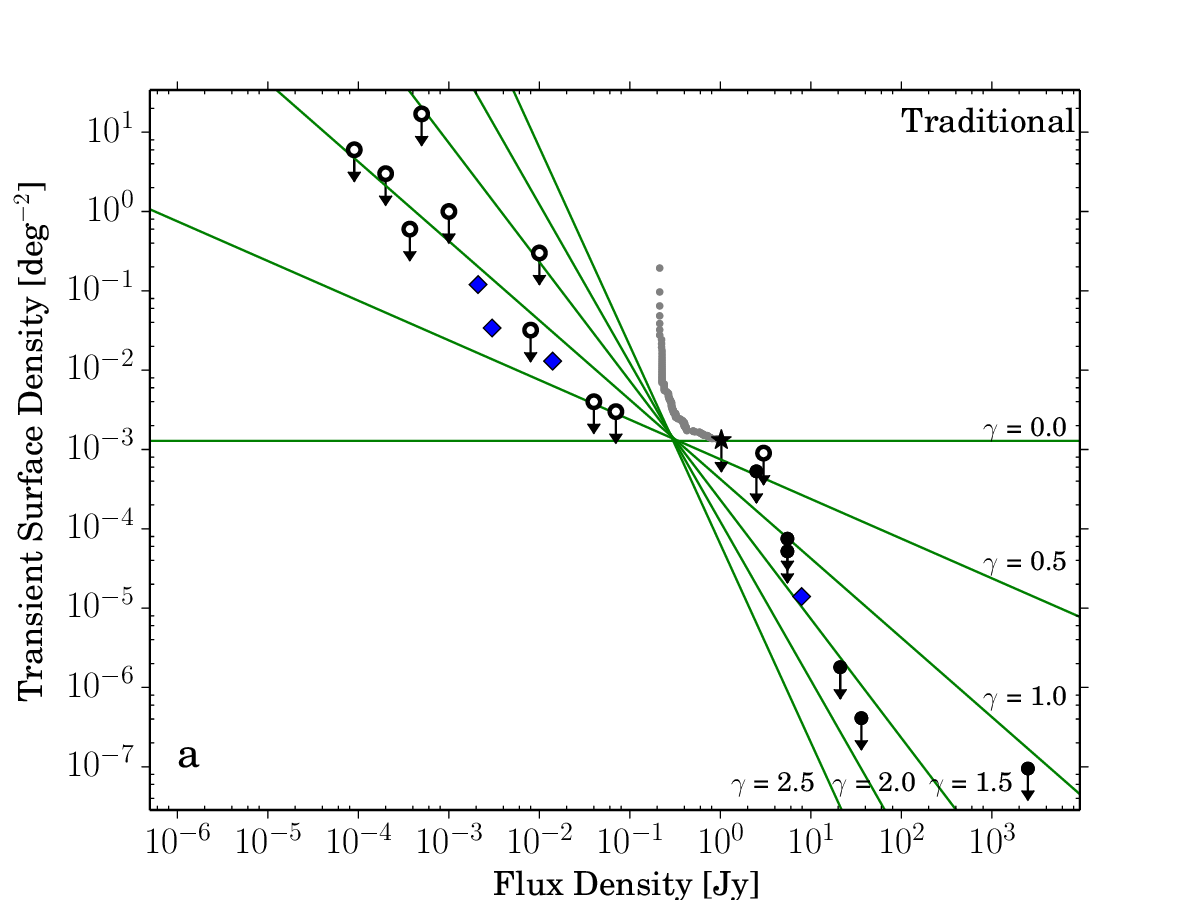}
\includegraphics[scale=0.45,viewport=67 0 522 396,clip]{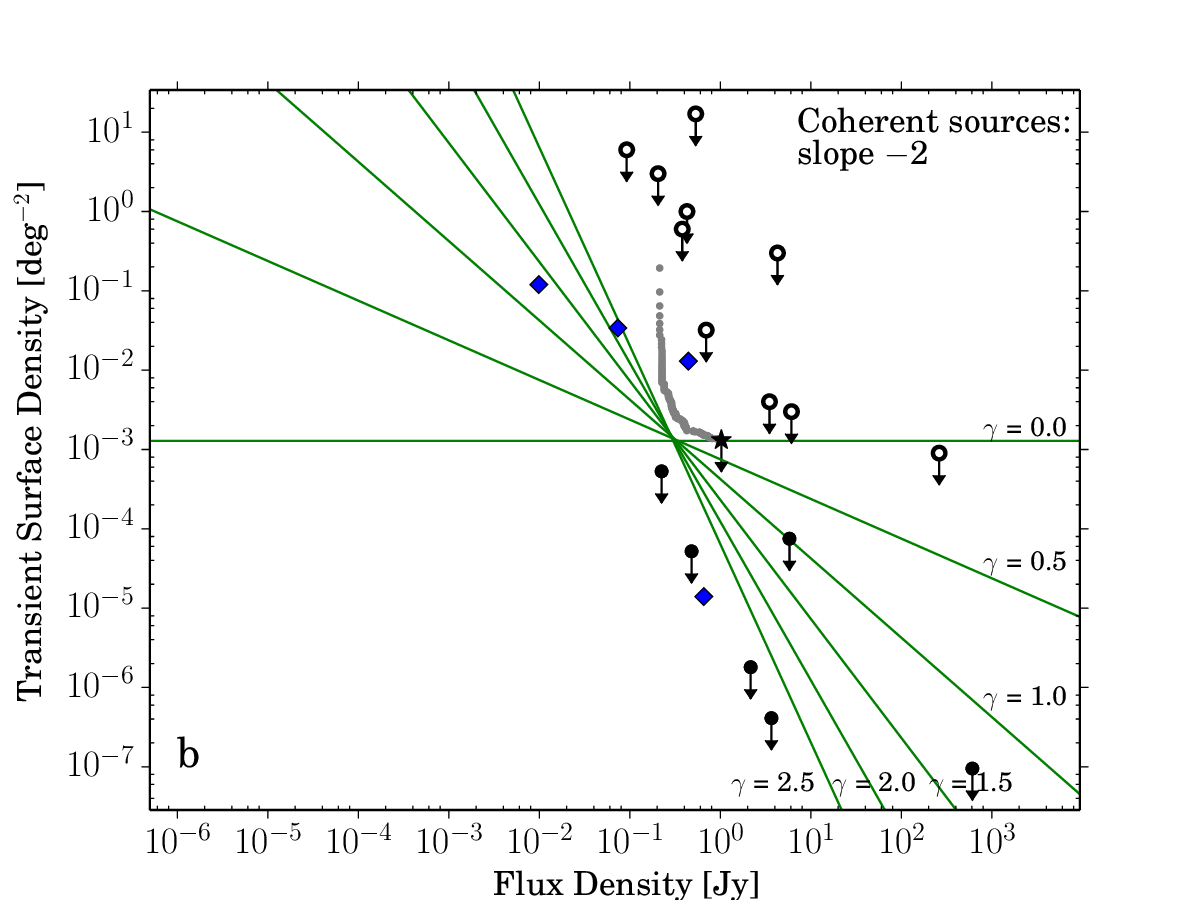}
\includegraphics[scale=0.45,viewport=0 0 522 396, clip]{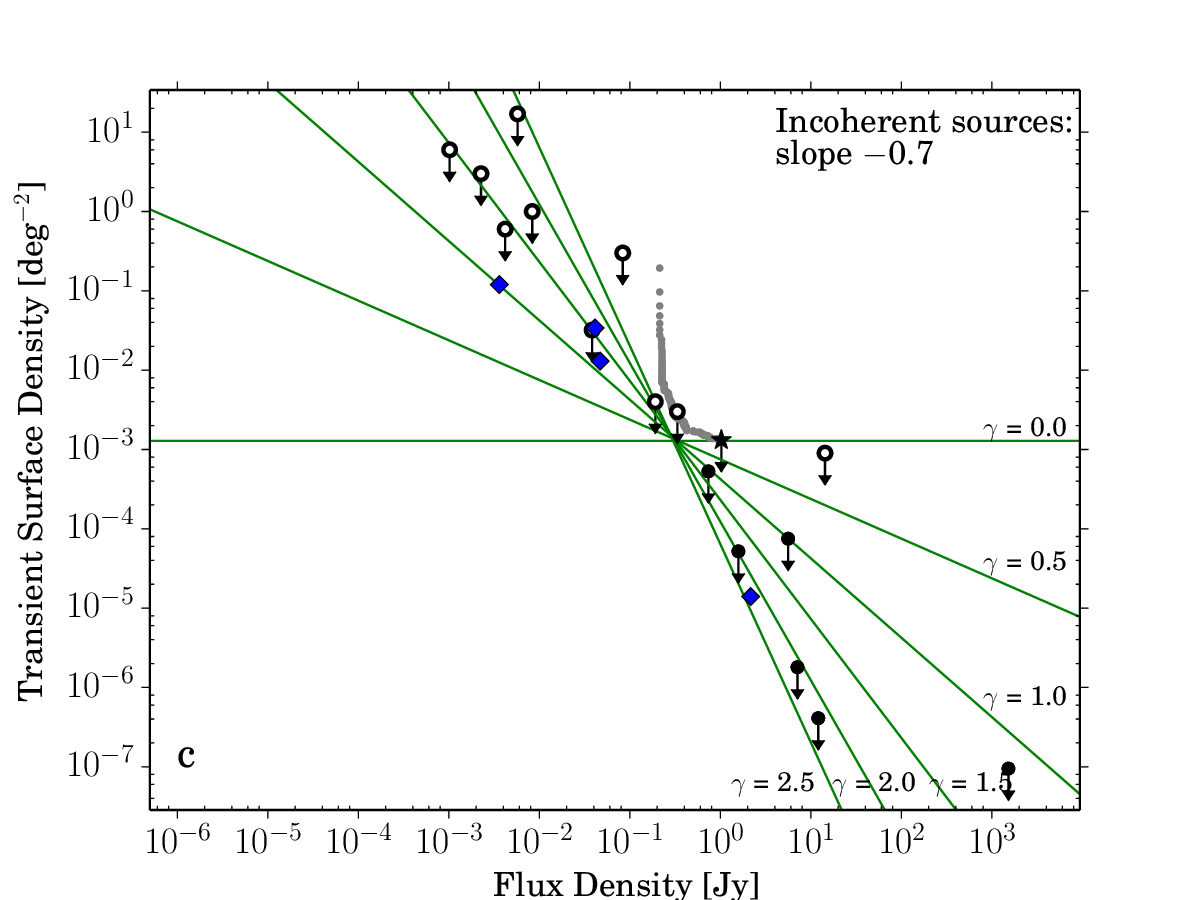}
\includegraphics[scale=0.45,viewport=67 0 522 396, clip]{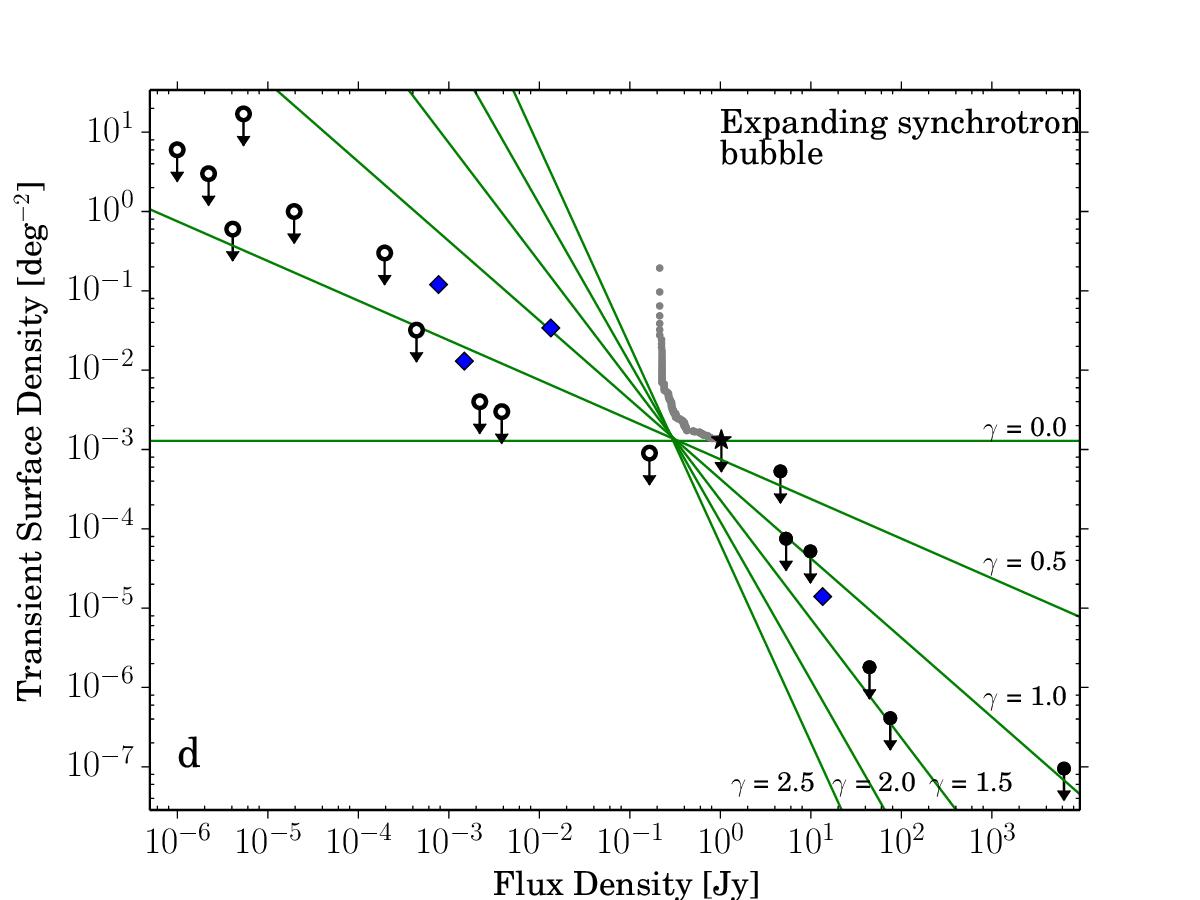}
\caption{Transient surface density $\left(\mbox{deg}^{-2}\right)$ against flux density (Jy) of detections of
transients (diamonds) and upper limits based on non detections (circles).
In these four plots we aim to compare our result from Figure~\ref{fig:methods} with that from other radio
transient surveys, despite the difficulty that these surveys have been done at rather different frequencies,
and the spectral indices of radio sources span a wide range.
The top left panel shows the
comparison as it is usually made in previous literature, ignoring the frequency difference between surveys.
In the other three, we have extrapolated to a common frequency of 150\,MHz. This requires using an assumption on the
source spectrum, so we show three different ones, which we hope span the expected range of spectral index.
In the top right panel the fluxes have been extrapolated assuming a spectral index of $-2$, typical for coherent radio
emitters, in the bottom left assuming a spectral index of $-0.7$ (optically thin synchrotron emission), while in the bottom
right assuming a van der Laan model for an expanding synchrotron bubble.
The star represents the upper limit from this work using the same technique as the other surveys in the literature.
The grey circles indicate the upper limits derived from this work
with a method free of assumptions on the flux distribution of transient sources. The lines indicate the upper
limits assuming a power-law distribution in flux for the transient sources as in
Equations~\ref{eq:flux distribution}-\ref{eq:k upper limit}.
The different lines represent different values of the exponent $\gamma$.
The green lines have been extended throughout the flux density axis for comparison purposes.
Open symbols represent surveys taken at frequencies above 1\,GHz. The surveys displayed here are listed in
Table~\ref{tab:comparison}.
The top right plot shows that our survey is indeed quite competitive with most previous ones for steep-spectrum sources,
but that, as expected,  surveys for incoherent synchrotron transients are better conducted at GHz frequencies.
}
\label{fig:compare}
\end{center}
\end{figure*}

Over the last decade, radio telescope archives, containing large volumes of data (particularly for calibrator fields),
have been searched for transients and variables.
An archival study comparing the NVSS \citep{NVSS_Condon1998} and FIRST \citep[Faint
Images of the Radio Sky at Twenty-cm;][]{FIRST_Becker1995} catalogues was conducted by
\citet{GRBorphan_Levinson2002}, with a follow-up study by \citet{radiooptfollowup_GalYam2006};
no transients were found. This survey is not reported in Figure~\ref{fig:compare} because its upper limit regards gamma-ray
burst afterglows only and is not representative of the whole radio transients population.
\citet{VLAarchivetrans_Bower2007} analysed 944 epochs of archival VLA data on the same field at 4.8 and
8.4\,GHz spanning a period of 22 years reporting the discovery of ten radio transients.
\citet{radiotransrevision_Frail2012} reanalysed this dataset and reported that more than half of these transients
were either caused by rare data reduction artefacts, or that the detections had a lower signal-to-noise ratio after re-reduction.
For our comparison we have adopted the conservative snapshot rate calculated by \citet{radiotransrevision_Frail2012},
which assumes no detections, even though they note that one source might have been a real transient.
\citet{MOSTtrans_Bannister2011} published results from a search for transient and variable sources in
the Molonglo Observatory Synthesis Telescope (MOST) archive at 843\,MHz; 15 transients and 53 highly-variable
sources were detected over a 22-year period.
\citet{VLAradiotrans_BowerSaul2010} have published further archival work examining observations of the calibrator
3C286 at 1.4\,GHz. They examined 1852 epochs covering over 23 years and no radio transients were reported.
\citet{VLAarchival_Bell2011} also reported results from an analysis of archival VLA data at 1.4, 4.8 and 8.4\,GHz of
several commonly observed calibrator fields covering 24 years, but no transients were detected.
\citet{SN&trans_Alexander2014} have recently reported on their monitoring campaign on SN1994I in M51. During
their observations they collected many hours of data using the VLA at 4.9\,GHz and searched it for radio transients but
did not find any.

At low radio frequencies ($<$ 500\,MHz) only a few blind searches for variable and transient sources have been performed.
\citet{Adam_Stewart2015} performed a survey around the North Celestial Pole at 60\,MHz with LOFAR and detected one
transient. They reported results on five different time-scales, from 30 seconds to 297 minutes.
\citet{trsearch_Lazio2010} conducted a survey for transients at 74\,MHz with the Long Wavelength
Demonstrator Array (LWDA). No transients were found in their dataset.
\citet[][]{FLIRT_Balsano1999}, using the Fallbrook Low-frequency Immediate Response Telescope (FLIRT), operating at
74\,MHz looked for prompt emission from gamma-ray bursts but did not find any brighter than 400\,Jy in their survey.
\citet{MWAtrans_Bell2014} performed a survey campaign with the MWA at 154\,MHz and reported no transient detections.
\citet{trsearch_Jaeger2012} conducted the deepest blind transient survey below 500\,MHz.
They used six epochs of Very Large Array observations at 325\,MHz centred on the \textit{Spitzer Space Telescope} Wide-field
InfraRed Extragalactic (SWIRE) deep field. They report the detection of one radio transient.

\citet{Rates_Trott2013} developed a framework to calculate the transient rate of beamformed data starting from technical
parameters of the survey and assuming a flux distribution for the astrophysical source population.
\citet[][]{SKA_memo_Cordes2007} adopted a more general approach to determine the number of transient sources of a
specified population that should be detected in a given survey starting from the properties of the transient population, as well
as the properties of the survey.
The relevant source properties are luminosity, duration of the transient, period, rate, and number density.
The properties of the observing campaign are sampled area, duration and speed of the observations, characteristics
of the instrument (such as noise and resolution).
From these parameters it is possible to estimate the probability that a transient source is detectable when it is observed,
and the total number of sources that should be detected.
\citet[][]{SKA_memo_Cordes2007} also derived the figure of merit of a transient survey which depends on bandwidth,
field of view, detection threshold,
number of transient sources being detectable at the same time, and the ratio between the telescope dwell time and
the duration of the transients.
These analyses therefore focus on the inverse of our problem, namely to predict how many sources of a known type
and rate a survey with certain properties will detect, i.e., an essentially model-dependent exercise. Our goal here
is to explore what population of yet unknown transients might lie hidden in an unexplored part of parameter space,
and we aim to state our limits as much as possible in a model-independent way. As inverses, the two processes are
of course somewhat related, and we will explore the comparison in further detail in a follow up study which is
currently in preparation.

Assuming that the three GCRT transients are indicative of a more general distributed population, then we would expect
a surface density $\sim$\,3\,/\,68.8\,deg$^{-2}$\,=\,4.4\,$\cdot\,10^{-2}$\,deg$^{-2}$ at a flux limit of 0.1\,Jy.
Converting this value to 0.5\,Jy we get surface densities between 2\,$\cdot\,10^{-2}$ and 8\,$\cdot\,10^{-4}$\,deg$^{-2}$,
which is close to the expected rate of GCRT type events. This survey therefore rules out a
significant bright coherent population of GCRT type transients.

\begin{table*}
\begin{center}
\begin{tabular}{|c|c|c|c|c|}
\hline
\hline
Survey				& Sensitivity (mJy)		& $\rho$ (deg$^{-2}$)	& t$_{\textrm{char}}$		& $\nu$ (GHz) 		\\
\hline
Stewart et al. 2015		& $>$ 22900			& $<$ 4.1 $\cdot 10^{-7}$	& 30 seconds			& 0.060			\\
Stewart et al. 2015		& $>$ 13500			& $<$ 1.8 $\cdot 10^{-6}$	& 2 minutes			& 0.060			\\
Stewart et al. 2015		& $>$ 3000			& $<$ 5.2 $\cdot 10^{-5}$	& 55 minutes			& 0.060			\\
Stewart et al. 2015		& $>$ 1400			& $<$ 5.3 $\cdot 10^{-4}$	& 297 minutes			& 0.060			\\
Lazio et al. 2010		& $>$ 2.5 $\cdot 10^6$	& $<$ 9.5 $\cdot 10^{-8}$	& 5 minutes			& 0.0738			\\
This work				& $>$ 500				& $<$ 0.001			& minutes -- months		& 0.150			\\
Bell et al. 2014			& $>$ 5500			& $<$ 7.5 $\cdot 10^{-5}$	& minutes -- year $^a$	& 0.154			\\
Bower \& Saul 2011		& $>$ 70				& $<$ 0.003			& 1 day				& 1.4				\\
Bower \& Saul 2011		& $>$ 3000			& $<$ 9 $\cdot 10^{-4}$	& 1 day				& 1.4				\\
Croft et al. 2010		& $>$ 40				& $<$ 0.004			& 81 days -- 15 years	& 1.4				\\
Gal-Yam et al. 2006 $^b$	& $>$ 6				& $<$ 1.5 $\cdot 10^{-3}$	& -					& 1.4				\\ 
Bell et al. 2011			& $>$ 8				& $<$ 0.032		 	& 4.3 -- 45.3 days		& 1.4, 4.8 and 8.4	\\
Bower et al 2010		& $>$ 1				& $<$ 1		 		& 1 month				& 3.1				\\
Bower et al 2010		& $>$ 10				& $<$ 0.3		 		& 1 month				& 3.1				\\
Frail et al. 2012			& $>$ 0.09			& $<$ 6		 		& 1 year				& 4.8 and 8.4		\\
Frail et al. 2012			& $>$ 0.37			& $<$ 0.6				& 20 minutes -- week	& 4.8 and 8.4		\\
Frail et al. 2012			& $>$ 0.20			& $<$ 3				& 2 months			& 4.8 and 8.4		\\
Alexander et al. 2014	& $>$ 0.5				& $<$ 17				& minutes -- months		& 4.9				\\
\hline
Stewart et al. 2015		& $>$ 4100			& 1.4 $\cdot 10^{-5}$	& 11 minutes			& 0.060			\\
Hyman et al. 2009		& $>$ 30				& 0.034 $^c$			& days -- months		& 0.235 and 0.330	\\
Jaeger et al. 2012		& $>$ 2.1				& 0.12				& 1 day -- 3 months		& 0.325			\\
Bannister et al. 2011		& $>$ 14				& 0.013				& days -- years			& 0.843			\\
\hline
\hline
\end{tabular}
\caption{Summary of the radio transient surface densities reported in the literature.
The table is split in two parts:
the upper part contains upper limits based on non detections while the bottom part shows transient detections.
The results from \citet{Adam_Stewart2015} and \citet{radiotransrevision_Frail2012} have been stated multiple times depending
on the characteristic time-scale sampled. \citet{PiGSS_Bower2010} and \citet{VLAradiotrans_BowerSaul2010} state two
different rates depending on the flux density.\newline
$^a$ The authors note that they are most sensitive to time-scales of 26 minutes and 1 year. This is displayed in
Figure~\ref{fig:compare_timescales} with two circles indicating these specific time-scales.\newline
$^b$ The results from this work regard gamma-ray burst afterglows only and are not representative for the whole
radio transients population. This data point is not displayed in Figure~\ref{fig:compare} and
Figure~\ref{fig:compare_timescales}.\newline
$^c$ The transient surface density for this survey is obtained from the calculation performed by \citet{ASGARD_Williams2013}
which takes into account results from  \citet{GCRT1746_Hyman2002, GCRT1745_Hyman2006, GCRT1742_Hyman2009}.}
\label{tab:comparison}
\end{center}
\end{table*}

In Figure~\ref{fig:compare} we compare the limits on the transient surface density from this study with those found
in the literature
(other works comparing transient surface densities are for example \citealt{VLAarchival_Bell2011,
radiotransrevision_Frail2012} and \citealt{SN&trans_Alexander2014}).
The surveys from Table~\ref{tab:comparison} are represented with circles in the case of non-detections
or with diamonds for the detections.
One should also take into account that these data points rely on few detections and their uncertainties
are not known.
Note that the constraints imposed by this work are comparable to those from other surveys.
Figure~\ref{fig:compare}a represents the transient surface density as a function of the
sensitivity as it is given in the literature, without any corrections due to the fact that surveys were performed
at different frequencies.

In Figure~\ref{fig:compare} we also extrapolated the flux limits calculated in the other surveys to the same
frequency as ours (150\,MHz) according to three different scenarios.
In panel b, we evaluated the case in which transient sources are emitting coherent radiation.
As explained in Section~\ref{sec:intro}, these sources are expected to have steep spectra, and we corrected their
flux densities assuming a spectral index of --2.
In panel c we analysed the case in which the transient sources are
optically thin synchrotron emitters, having a radio spectrum characterised by a simple power law with the
spectral index depending on the exponent of the energy distribution of the electrons responsible for the
synchrotron emission ({\it p}),

\begin{equation}
S (\nu_{2}) = S(\nu_{1}) \left(\frac{\nu_{2}}{\nu_{1}}\right)^{-(p - 1)/2} .
\label{eq:optically thin synchrotron}
\end{equation}

In panel d of Figure~\ref{fig:compare} we show the case in which transients are non-relativistically
expanding synchrotron bubbles \citep[][]{timescale_frequency_VanDerLaan1966}. In this case their spectra
show a peak that is shifting towards lower frequencies and declining as the source evolves; this is the typical
scenario for a radio source resulting from an astrophysical explosion like a supernova or a gamma-ray burst.
Since almost all surveys shown in Figure~\ref{fig:compare} resulted in non-detections, we treated them as upper
limits on the peak flux  of the sources. 
We extrapolated all the upper limits to the frequency of our survey using the relation for the peak flux given in
\citet{timescale_frequency_VanDerLaan1966}, which also depends on {\it p},

\begin{equation}
S_{\mathrm{max}} (\nu_{2}) = S_{\mathrm{max}}(\nu_{1}) \left(\frac{\nu_{2}}{\nu_{1}}\right)^{(7p + 3)/(4p + 6)} .
\label{eq:vdL model}
\end{equation}

For the detections in Figure~\ref{fig:compare} we used the same extrapolation for consistency.
Measurements of the spectra of optically thin synchrotron emitting sources showed that their
slopes are around $-0.7$, implying a value for {\it p} around 2.4 \citep[e.g.,][]{spectralindex_Kellermann1964}.
For consistency, this value was used in the non-relativistic synchrotron bubble scenario as well, yielding an `effective'
slope of +1.3 according to Equation~\ref{eq:vdL model}.

As previously mentioned, very few surveys have been performed at frequencies below 500 MHz.
The ones listed in Table~\ref{tab:comparison} have been plotted with filled symbols in
Figure~\ref{fig:compare}.

As we mentioned in Section~\ref{sec:intro}, low frequency surveys are most sensitive to coherently emitting sources.
This is clearly shown in Figure~\ref{fig:compare}b, where our survey is far more sensitive than all the ones
performed in the GHz regime, and the only other surveys that are as sensitive as ours were also performed at low frequencies.
In the case of optically thin synchrotron emitting sources displayed in Figure~\ref{fig:compare}c, surveys performed
to date at low and high radio frequencies seem to give comparable results, whereas in the case of an expanding synchrotron
bubble (Figure~\ref{fig:compare}d) high frequency radio surveys give better results.
It is also clear from Figure~\ref{fig:compare}b that a single power-law flux distribution cannot explain all the detections
and upper limits. This means either that the overall flux
distribution is not a power law or that the dominant population of transients does not have a power-law distribution over
the approximately 5 decades of source flux probed by the surveys to date, or that multiple populations of transients contribute.

\section{Conclusions}
\label{sec:conclusions}
We have presented the results from a radio transient search at very low frequencies with LOFAR on four fields among the
Medium Deep fields monitored by the PanSTARRS consortium, and found no credible transients.

Regarding populations of transient sources at these low frequencies, we conclude that:
\begin{enumerate}
 \item We set a robust limit to the instantaneous surface density $\rho$ of transient sources lasting 15 min:
$\rho$(S\,$>$\,0.3\,Jy)\,$<$\,1.3\,$\cdot$\,10$^{-3}$\,deg$^{-2}$.
\item For longer transients, our surface density limit becomes poorer, gradually increasing to about 0.1\,deg$^{-2}$ at
100 days.
\item These limits are significantly stronger than previous, higher-frequency constraints for coherent, steep-spectrum sources,
comparable to those studies for optically thin synchrotron sources, and weaker for Van der Laan-type synchrotron bubble
outbursts.
\item Our limits on the rates of transients are more constraining for longer events, since these fall in between observations less
easily, and range from $\sim$\,100 sky$^{-1}$ day$^{-1}$ to $\sim$\,10 sky$^{-1}$ day$^{-1}$ for time-scales from 15 min to 100 days.
\item We show that a single power-law source count distribution and power-law spectral shape of transients cannot explain the
combination of detections and limits of existing surveys, and that our survey must be fairly close (to) the size needed to start detecting
transients.
\end{enumerate}

Regarding methods to constrain the surface density and rate of transients from a given survey, we find that:
\begin{enumerate}
\item The traditional method of setting a surface density limit can be improved upon robustly by a factor 3--4 in flux limit and
density limit.
\item We present a new method to derive model-independent constraints to the transient surface density that consists of
iteratively discarding the noisiest image from the series and finding an optimum combination of area covered and flux limit set
around the modal sensitivity of the collection of images. It improves the effective flux limit of the survey by a factor 3 relative to
the traditional method.
\item We present a yet stronger new method of setting a surface density limit, which is initially model dependent in that it
assumes the source count distribution is a power law over the range of fluxes covered by the image flux limits, but eventually
is virtually independent of that assumption. It improves the rate limit one can set by about a factor 4 relative to the previous
method.
\item The traditional method of converting the surface density limit to a rate limit is incorrect: it ignores the strong duration
dependence of the limit one can set, and for transient of long duration (weeks to months) gives a limit that is 10--100 times
too weak.
\item We derive an approximate analytic method by which one can correctly convert the surface density limit into a rate limit,
for any specific survey, as a function of transient duration.
\end{enumerate}

Lastly, we note that in this survey we have used only a (relatively) small amount of data relative to the full data volumes that are
becoming available in large LOFAR surveys. Also, the techniques for calibrating and imaging wide-field, low-frequency radio
interferometric data are still improving significantly, so the per-image sensitivity will also still improve. As a result, much more
powerful low-frequency transient surveys will soon occur.

\section*{Acknowledgments}
The authors would like to thank the referee for the useful comments and suggestions that helped to improve the paper.
The authors would like to thank the LOFAR observatory staff for their assistance in obtaining and the handling
of this large data set.
LOFAR, the LOw-Frequency ARray designed and constructed by ASTRON, has facilities in several countries,
that are owned by various parties (each with their own funding sources), and that are collectively operated by
the International LOFAR Telescope (ILT) foundation under a joint scientific policy.
DC, AJvdH, RAMJW, JDS, AR, YC acknowledge support from the European Research Council Advanced Grant
247295 ``AARTFAAC'' (P.I. R.~A.~M.~J.~Wijers).
JWB,  AJS, RPF acknowledge support from the European Research Council Advanced Grant 267697 ``4 Pi Sky''
(P.I. R. Fender). JWTH acknowledges funding from an NWO Vidi fellowship and European Research
Council Starting Grant 337062 ``DRAGNET''.
SC acknowledges financial support from the UnivEarthS Labex program of Sorbonne Paris Cit\'e (ANR-10-LABX-0023
and ANR-11-IDEX-0005-02).

\bibliographystyle{mnras.bst}
\bibliography{/Users/dariocarbone/Documents/carbone_bibliography.bib}

\end{document}